\input harvmac
\noblackbox

\def\appA{A}
\def\appB{B}
\def\appC{C}
\def\appD{D}

\def\h {{1\over 2}}
\def\al {\alpha}

\def\ov {\overline}
\def\o {\over}
\def\IZ{ {\bf Z}}
\def\IF{{\bf F}}
\def\IR{ {\bf R}}

\def\cmp {{\it Comm. Math. Phys.\ }}
\def\np {{\it Nucl. Phys.\ } {\bf B\ }}
\def\pl {{\it Phys. Lett.\ } {\bf B\ }}
\def\pr {{\it Phys. Rept.\ } }

\def\br{\hfill\break}

\def\cB {\cal B}
\def\Li {{\cal L}i}
\def\Zc {{\cal Z}}

\def\Cc {\cal C}

\def\th {\theta}

\def\Ga {\Gamma}
\def\si {\sigma}
\def\Bc {\cal B}
\def\tr {{\rm tr}}

\def\mod {{\rm mod}}
\def\lf {\left}
\def\ri {\right}
\def\ra {\rightarrow}
\def\lra {\longrightarrow}
\def\al {\alpha}
\def\re {{\rm Re}}
\def\im {{\rm Im}}
\def\p {\partial}

\lref\cf{P. Candelas and A. Font, \np {\bf 511} (1998) 295;\br
G. Aldazabal, A. Font, L.E. Ib{\'a}{\~n}ez and A.M. Uranga,
\np {\bf 492} (1997) 119.}

\lref\kawa{Y. Katsuki, Y. Kawamura, T. Kobayashi, N. Ohtsubo, Y. Ono
and K. Tanioka, {\it Tables of $\IZ_N$ orbifold models}, Kanazawa DPKU--8904.}

\lref\solving{E. Kiritsis, C. Kounnas, P.M. Petropoulos and J. Rizos
\pl {\bf 385} (1996) 87.}
\lref\benakli{K. Benakli, hep--th/9805181.}
\lref\choi{K. Choi, {\it Phys. Rev.} {\bf D 56} (1997) 6588.}

\lref\FT{S. Ferrara and S. Theisen, 
{\it Moduli spaces, effective actions and 
duality symmetry in string compactifications}, CERN-TH-5652-90, 3rd Hellenic
Summer School, Corfu, Greece, Sep 13-23, 1989.} 

\lref\silver{S. Kachru and E. Silverstein, \np {\bf 504} (1997) 272.}
\lref\GSW{M.B. Green, J.H. Schwarz and E. Witten, {\it Superstring
theory}, Volume 2, Cambridge 1987.}
\lref\cecotti{M. Bershadsky, S. Cecotti, H. Ooguri and C. Vafa, 
\np {\bf 405} (1993) 279; \cmp {\bf 165} (1994) 311.}
\lref\hw{P. Horava and E. Witten, \np {\bf 460} (1996) 506;
             \np {\bf 475} (1996) 94.}
\lref\bd{T. Banks and M. Dine, \np {\bf 479} (1996) 173.}
\lref\ib{J.--P. Derendinger, L.E. Ib{\'a}{\~n}ez and H.P. Nilles, 
\np {\bf 267} (1986) 365;\br 
L.E. Ib{\'a}{\~n}ez and H.P. Nilles, \pl {\bf 169} (1986) 354;\br
K. Choi and J.E. Kim, \pl {\bf 165} (1985) 71.}

\lref\orbifolds{L. Dixon, J.A. Harvey, C. Vafa and E. Witten, 
\np {\bf 261} (1985) 678;\br \np {\bf 274 }(1986) 285;\br
L.E. Ib{\'a}{\~n}ez, J. Mas, H.P. Nilles and F. Quevedo, \np {\bf 301}
(1988) 157.}

\lref\koko{C. Kokorelis, hep--th/9802099.}

\lref\wend{{\it See e.g.:} 
P. Mayr and S. Stieberger, {\it Low energy properties
of (0,2) compactifications}, hep--th/9412196, 28th International Symposium
on Particle Theory, Wendisch--Rietz, Germany, 30 Aug - 3 Sep 1994.}

\lref\wyll{N. Wyllard, {\it JHEP} {\bf 4} (1998) 009.} 
\lref\fs{K.\ F{\"o}rger and S.\ Stieberger, \np {\bf 514} (1998) 135.}

\lref\wstrong{E. Witten, \np {\bf 471} (1996) 135.}
\lref\witten{E. Witten, \np {\bf 460} (1996) 541.}

\lref\gsw{M. Green, J.H. Schwarz and P. West, \np {\bf 254} (1985) 327.}
\lref\sw{N. Seiberg and E. Witten, \np {\bf 471} (1996) 121.}
\lref\sch{J.H. Schwarz, \pl {\bf 371} (1996) 223.}
\lref\schall{M. Berkooz, R.G. Leigh, J. Polchinski, J.H. Schwarz, 
N. Seiberg and E. Witten, \np {\bf 475} (1996) 115.}
\lref\erler{J. Erler, {\it J. Math. Phys. } {\bf 35} (1994) 1819.}
\lref\dmw{M. Duff, R. Minasian and E. Witten, \np {\bf 465} (1996) 413.}

\lref\cand{T. Eguchi, P.B. Gilkey and  A.J. Hanson, {\it Phys. Rept.}
{\bf 66} (1980) 213.}
\lref\ls{W. Lerche and S. Stieberger, {\it Prepotential, mirror map
and $F$--theory on $K3$}, hep--th/9804176.}
\lref\intri{K. Intriligator, \np {\bf 496} (1997) 177;\br
J.D. Blum and K. Intriligator, \np {\bf 506} (1997) 199.}
\lref\kmv{A. Klemm, P. Mayr and C. Vafa,  hep--th/9607013;\br
W. Lerche, P. Mayr and N.P. Warner, \np {\bf 499} (1997) 125.}
\lref\ganor{O.J. Ganor and A. Hanany, \np {\bf 474} (1996) 122;\br
O.J. Ganor, \np {\bf 488} (1997) 223;\br
O.J. Ganor, D.R. Morrison and N. Seiberg, \np {\bf 487} (1997) 93.}

\lref\japan{T. Eguchi, H. Ooguri, A. Taormina and S. Yang, \np {\bf
315} (1989) 193;\br
T. Kawai, Y. Yamada and S. Yang, \np {\bf 414} (1994) 191.}
\lref\zagier{M. Eichler and D. Zagier, {\it The theory of Jacobi
forms}, Birkh{\"a}user 1985.}
\lref\ellg {A.\ Schellekens and N.\ Warner,
\np {\bf 287} (1987) 317;\br
E.\ Witten, \cmp {\bf 109} (1987) 525;\br
W. Lerche, B.E.W. Nilsson, A.N. Schellekens and N.P. Warner,
\np {\bf 299} (1988) 91;\br
W. Lerche, \np {\bf 308} (1988) 102;\br
W. Lerche, B.E.W. Nilsson and A.N. Schellekens, \np {\bf 289}
(1987) 609;\br 
W. Lerche, A.N. Schellekens and N.P. Warner, \pr {\bf 177} (1989) 1.}

\lref\per{P. Berglund, S. Katz, A. Klemm and P. Mayr, 
\np {\bf 483} (1997) 209.}

\lref\klF{V. Kaplunovsky and J. Louis, \pl {\bf 417} (1998) 45.}
\lref\cclm{G.L. Cardoso, G. Curio, D. L{\"u}st and T. Mohaupt, 
\pl {\bf 389} (1996) 479}
\lref\am{P. Aspinwall and D.R. Morrison, hep--th/9705104}
\lref\NG{V.A. Gritsenko and V.A. Nikulin, alg--geo/9611028.}
\lref\AL{P. Aspinwall and J. Louis, \pl {\bf 369} (1996) 233}
\lref\hmR{J.A. Harvey and G. Moore, hep--th/9611176.}
\lref\msf {P. Mayr and S. Stieberger, \pl {\bf 355} (1995) 107.}
\lref\agntz {I. Antoniadis, E. Gava, K. Narain and T. Taylor,
\np {\bf 432} (1994) 187.}
\lref\vk {V. Kaplunovsky, \np {\bf 307} (1988) 145 and hep--th/920570.}
\lref\dklz {L. Dixon, V. Kaplunovsky and J. Louis, \np {\bf 355} (1991) 649.}
\lref\ich{S. Stieberger, One--loop corrections and gauge coupling unification 
in superstring theory, Ph.D. thesis Munich May 1995, TUM--HEP--220/95.}
\lref\mm {M. Henningson and G. Moore, \np {\bf 482} (1996) 187.}
\lref\hm {J.A. Harvey and G. Moore, \np {\bf 463} (1996) 315.}
\lref\ke {T. Kawai, \pl {\bf 372} (1996) 59.}
\lref\kz {T. Kawai, \pl {\bf 397} (1997) 51.}
\lref\ccl {G.L. Cardoso, G. Curio, and D. L{\"u}st, \np {\bf 491} (1997) 147.}
\lref\car {G.L. Cardoso, hep--th/9612200.}
\lref\dewit {I. Antoniadis, S. Ferrara, E. Gava, K. Narain and T. Taylor,
\np {\bf 447} (1995) 35;\br 
B. de Wit, V. Kaplunovsky, J. Louis and D. L{\"u}st, 
\np {\bf 451} (1995) 53.}

\lref\janvadim{V. Kaplunovsky and J. Louis, \np {\bf 444} (1995) 191.}
\lref\nsz {H.P. Nilles and S. Stieberger, \np {\bf 499} (1997) 3.}

\lref\borch {R.E. Borcherds, Invent. Math. {\bf 120} (1995) 161.}
\lref\sonne {J. Louis, J. Sonnenschein, S. Theisen and S. Yankielowicz,
\np {\bf 480} (1996) 185.}
\lref\dfkz {J.P. Derendinger, S. Ferrara, C. Kounnas and F. Zwirner, 
\np {\bf 372} (1992) 145;\br
I. Antoniadis, E. Gava, K.S. Narain and T.R. Taylor,
\np {\bf 407} (1993) 706.}

\lref\hpn {L.E. Ib{\'a}{\~n}ez, H.P. Nilles and F. Quevedo,
\pl {\bf 187} (1987) 25; \pl {\bf 192} (1987) 332}
\lref\dhvw {L. Dixon, J. Harvey, C. Vafa and E. Witten,
\np {\bf 261} (1985) 678; {\bf B 274} (1986) 285;
L.E. Ib{\'a}{\~n}ez, J. Mas, H.P. Nilles and F. Quevedo,
\np {\bf 301} (1988) 157}

\lref\kk{E. Kiritsis and  C. Kounnas, 
\np {\bf 442} (1995) 472; {\it Nucl. Phys. Proc. Suppl.} {\bf 45BC}
(1996) 207;\br
P.M. Petropoulos and J. Rizos, {Phys. Lett.} {\bf B 374} (1996) 49.}

\lref\kkpr{E. Kiritsis, C. Kounnas, P.M. Petropoulos and J. Rizos,
\np {\bf 483} (1997) 141.}

\lref\afiq {G. Aldazabal, A. Font, L.E. Ib{\'a}{\~n}ez and F. Quevedo, 
\np {\bf 461} (1996) 85.}

\lref\afiuv{G. Aldazabal, A. Font, L.E. Ib{\'a}{\~n}ez, A.M. Uranga and
G. Violero, \np {\bf 519} (1998) 239.}

\lref\mse {P. Mayr and S. Stieberger, \np {\bf 407} (1993) 725;\br
D. Bailin, A. Love, W. Sabra and S. Thomas, 
{Mod. Phys. Lett.} {\bf A9} (1994) 67; {\bf A10} (1995) 337.}

\lref\kv{S. Kachru and C. Vafa, \np {\bf 450} (1995) 69.}
\lref\klm{A. Klemm, W. Lerche and P. Mayr, \pl {\bf 357} (1995) 313.}
\lref\byau{B. Lian and S.-T. Yau, hep--th/9507151, hep--th/9507153.}
\lref\kklmv{S. Kachru, A. Klemm, W. Lerche, P. Mayr and C. Vafa,
\np {\bf 459} (1996) 537.}
\lref\anpar{I. Antoniadis and H. Partouche, \np {\bf 460} (1996) 470.}
\lref\candel{P. Candelas, X. de la Ossa, A. Font, S. Katz and D. Morrison,
\np {\bf 416} (1994) 481;\br
S. Hosono, A. Klemm, S. Theisen and S.T. Yau, \cmp {\bf 167} (1995) 301;
\np {\bf 433} (1995) 501.}

\lref\mans{M. Henningson and G. Moore, \np {\bf 472} (1996) 518;\br
P. Berglund, M. Henningson and N. Wyllard,  {\it Nucl.Phys.} {\bf B 503}
(1997) 256.}

\lref\hpn{H.P. Nilles, M. Olechowski and M. Yamaguchi, {\it Phys. Lett.}
             {\bf B 415} (1997) 24; hep-th/9801030.}
\lref\andre{A. Lukas, B. A. Ovrut and D. Waldram, hep-th/9801087,
{\it Phys. Rev.} {\bf D 57} (1998) 7529, hep--th/9710208, hep-th/9709214,
hep--th/9803235;\br 
A. Lukas, B. A. Ovrut, K.S. Stelle and D. Waldram, hep--th/9806051.}
\lref\all{
I. Antoniadis and M. Quir{\'o}s, {\it Phys. Lett. } {\bf B 392} (1997) 61;\br
T. Li, J.~L. Lopez and D.~V. Nanopoulos, {\it Mod. Phys. Lett.}
{\bf A12} (1997) 2647;\br
E. Dudas and C. Grojean, {\it Nucl. Phys.} {\bf B 507} (1997) 553;\br
E. Dudas and J. Mourad, {\it Phys. Lett.} {\bf B400} (1997) 71;\br
I. Antoniadis and M. Quir{\'o}s, {\it Phys. Lett.} {\bf B 416} (1998) 327;\br
Z. Lalak and S. Thomas, \np {\bf 515} (1998) 55;\br
E. Dudas, \pl {\bf 416} (1998) 309;\br
J. Ellis, A.~E. Faraggi and D.~V. Nanopoulos, \pl {\bf 419} (1998) 123;\br
K. Choi, H.~B. Kim and C. Mu{\~n}oz, hep-th/9711158;\br
T. Li, hep-th/9801123;\br
M. Faux, hep-th/9801204;\br
P. Majumdar and S. SenGupta, hep-th/9802111;\br
D. Bailin, G.V. Kraniotis and A. Love, hep-ph/9803274;\br
J. Ellis, Z. Lalak, S. Pokorski and W. Pokorski, hep-ph/9805377.}

\lref\allk{P. Aspinwall, hep--th/9611137;\br
A. Klemm, hep--th/9705131;\br
L.E. Ib{\'a}{\~n}ez and A.M. Uranga, hep--th/9707075;\br 
D. L{\"u}st, hep--th/9803072.}

\Title{\vbox{\rightline{\tt hep-th/9807124} 
\rightline{CERN--TH/98--228}
\rightline{NEIP--001/97}}}
{\vbox{\centerline{$(0,2)$ Heterotic Gauge Couplings } 
\vskip4pt\centerline{and their $M$--Theory Origin}}}

\centerline{S. Stieberger}

\bigskip\centerline{\it CERN, Theory Division} 
\centerline{\it CH--1211 Gen{\`e}ve 23, SWITZERLAND}

\vskip .5in

We work out the relation between automorphic forms on $SO(2+s,2,\IZ)$
and gauge one--loop corrections of heterotic $K3\times T^2$ string 
compactifications for the cases $s=0,1$. We find that
one--loop gauge corrections of any orbifold limit of $K3$ 
can always be 
expressed by their instanton numbers and generic automorphic forms. 
These functions classify also one--loop gauge thresholds  of N=1 
$(0,2)$ heterotic compactifications based on toroidal orbifolds $T^6/\IZ_\nu$.
We compare these results with the gauge couplings of $M$--theory
compactified on $S^1/\IZ_2\times T^6/\IZ_\nu$ using 
Witten's Calabi--Yau strong coupling expansion.

\Date{07/98} 

\newsec{Introduction}

Within the last years, many 
perturbative calculations have been accumulated in heterotic $K3\times T^2$
compactifications, whose scalar field sector of N=2 vector multiplets
contains, besides the generic
$S,T,U$--moduli, which describe the dilaton, the size and shape of the
torus $T^2$, respectively, in addition Wilson line moduli. 
The latter parametrize non--trivial gauge background fields w.r.t. to the
internal torus $T^2$. These results --summarized for the two derivative 
couplings  in a holomorphic prepotential ${\cal H}$ and two 
other functions-- lead to a contr{\^o}l of all  
perturbative one--loop corrections in the gauge sector
\msf\ke\hm\mm\ich, in the gravitational sector \hm\mm\kz\ccl\car\
and the one--loop K{\"a}hler corrections via the prepotential \dewit\fs\koko.
N=2 heterotic--type-II duality \kv\ links the heterotic prepotential,
given as sum over trilogarithms \hm\ with the type-II prepotential, 
given as weighted instanton expansions \candel.  
The main evidence for the equivalence of a pair with a rank three gauge group 
was the appearance of the $j$--function, the automorphic
function of the perturbative duality group of the heterotic side, in the
functional dependence of the CY couplings at a certain boundary 
of the CY moduli space, which has been identified with 
the weak coupling limit of the heterotic string. The appearance
of automorphic functions of subgroups of $SL(2,\IZ)$, the typical 
dependence of the perturbative heterotic couplings, is a general 
phenomenon in CY spaces of a special fibration structure which has been
realized  in \klm. Moreover it was demonstrated there, that this
K3 fibration structure\foot{
Some articles reporting $K3$ dynamics in the context of
string--duality are the refs. \allk.} 
implies the appearance of automorphic functions
of modular groups of more variables, a mathematically surprising fact 
which was subsequently explored in \byau. In this way type-II
heterotic duality imposes surprising relations of CY mirror maps
to automorphic functions of heterotic duality groups, e.g. $SO(2+s,2,\IZ)$. 
Usually on the heterotic side, couplings are calculated as power series
in $exp(2\pi iS)$. These powers control the non--perturbative contributions
coming from space--time instantons and their coefficients themselves are 
automorphic functions under the perturbative duality group $SO(2+s,2,\IZ)$ 
including exchange symmetries \byau\mans. 
However, if the type-II heterotic duality provides information about 
CY periods in the perturbative heterotic regime, the opposite can be said 
about the strongly coupled phase. In \klm\kklmv\anpar\ heterotic--type-II
duality was used to derive the non--perturbative duality group mixing the 
dilaton with the other moduli. In particular in this way one obtains the
generalized automorphic functions which reduce to the perturbative heterotic 
ones at a special boundary of the CY moduli space.
This limit corresponds to the large base--space limit of the K3--fibration 
on the type-II side \AL.
In this limit the automorphic functions on the heterotic side
carry the information about the K3--fiber of the type-II side, i.e. in 
particular the type of singularity on the K3.
The simplest cases are the two or three moduli cases 
with the perturbative duality groups $SO(2,2,\IZ)$ and $SO(3,2,\IZ)$, 
respectively. Generic formulae may also be given for the higher groups
$SO(2+s,2,\IZ)\ ,\ s\geq 2$.
Whereas these symmetries often also appear in N=1 string vacua (see
e.g. \FT), not much
is known about the underlying modular functions, which unify N=1 amplitudes
in a similar way as described above for the N=2 case.
It is one aim of this paper to work out such a correspondence  for the gauge 
couplings of a class of N=1 string vacua. Due to the similarity
of the N=1 and N=2 target space duality groups and their underlying modular 
functions one might even guess, that again the relevant physics can 
be traced back to $K3\times T^2$ dynamics. 
Therefore, the aim of this paper is twofold:
We find a quite extensive and unifying description of N=2, $d=4$ 
gauge threshold 
corrections in terms of basic modular functions and spectrum dependent 
quantities.  
Second, we derive general expressions --given as $SO(2+s,2,\IZ)$ modular 
functions-- for the one--loop corrections to the
gauge couplings in N=1, $d=4$ theories with (0,2) superconformal symmetry on 
the world--sheet, realized as toroidal orbifolds. 
These are singular limits of Calabi--Yau manifolds (CYMs). 
Although this represents a specialization to a certain class
of string compactifications, this limit allows us to 
extract concrete results about (0,2) compactifications and
the full one--loop gauge couplings, including gauge group 
dependent and independent contributions. In contrast, for a smooth (2,2) 
CYM only the difference of the $E_6$ and $E_8'$ one--loop corrections
are known (given by the topological index).
The spectrum of toroidal orbifolds with N=1 space--time supersymmetry 
can be arranged into N=1, N=2 and N=4 multiplets, respectively, depending
on how their field representations are twisted along the world--sheet torus.
However, it is only the N=2 part, which gives rise to moduli dependent
perturbative gauge corrections.
Therefore, all the moduli dependence is encoded in $K3 \times T^2$ dynamics
and our study of $K3\times T^2$ gauge couplings may be used, to classify 
the N=1 couplings. In particular this means, that these couplings may be 
expressed by generic modular functions together with the instanton numbers of 
the underlying $K3 \times T^2$ compactifications, which represent 
N=2 subsectors of
the full N=1 orbifold. Our main results are presented in eqs. (4.4) and 
(4.5). This description allows us to recover the topological nature of these 
couplings. In particular, we are able to trace back their origin to 
Green--Schwarz terms in ten dimensions or gauge couplings of eleven 
dimensional $M$--theory. We give this link for N=1 orbifolds with both 
(2,2) and (0,2) superconformal world--sheet symmetry. This link provides also 
generic methods for the dimensional reduction on orbifolds, which is
useful for further investigations in $M$--theory phenomenology.

\newsec{Gauge and gravitational one--loop corrections in heterotic
$K3\times T^2$ compactifications}

We consider heterotic $K3\times T^2$ compactifications, where the instantons 
are embedded into $H\times H'$ subgroups\foot{Also combinations $U(1)\times H$
of Abelian and non--Abelian backgrounds are possible. In that case, 
the gauge group is
of the form $U(1)\times G$ \gsw.} of $E_8\times E_8$. 
Their instanton numbers, fulfill $n^{(1)}+n^{(2)}=\chi_{K3}\ ,\ 
\chi_{K3}=24$.
Therefore we define $n^{(1)}:=12-n, n^{(2)}:=12+n$ referring to $H,H'$, 
respectively.
This expresses the well--known fact, that for K3 compactifications, 
the instanton numbers have to add up to $24$, following from  
$\int_{K3}dH=0$, which guarantees a global well--defined $3$--form $H$ on $K3$
and the Bianchi identity $dH=\tr R^2-v_a\tr F_a^2$.
The remaining unbroken gauge group (the commutant $G^{(')}$ of $H^{(')}$)
is denoted by $G\times G'$.
In general, such vacua have $(0,4)$ world supersymmetry and N=2 space--time 
supersymmetry in $d=4$. 
The (new) supersymmetric index ${\cal Z}(q,\ov q)$ and variants of it
are the basic objects for 
string--amplitudes, which are obtained from it after taking the relevant 
order in the fields $R$ and $F$ and integrating over its modular 
invariant part \ellg.
For the case of vanishing Wilson lines $(s=0)$ and adjoint scalars
the (new) supersymmetric index $\Zc(q,\ov q)$ 
factorizes ($q=e^{2\pi i\tau}$)
\eqn\genus{\kern-1.5em\eqalign{
\Zc(q,\ov q)&=\eta^{-2}(q)
\tr_R\lf[q^{L_0-{c\o24}}\ov q^{\ov L_0-{\ov c\o24}}(F_L+F_R)
e^{\pi i(F_L+F_R)}\ri]_{(c,\ov c)=(22,9)}=Z_{K3}(q)Z_{2,2}(q,\ov q)\ ,\cr
&Z_{2,2}(q,\ov q,T,U)=\sum_{m_i,n^i} 
e^{2\pi i\tau(m_1n^1+m_2n^2)}\ 
e^{-{\pi\tau_2\o T_2U_2}}|TUn^2+Tn^1-Um_1+m_2|^2\ ,\cr}}
into a holomorphic $K3$--part $Z_{K3}(q)$ and a generic lattice sum 
$Z_{2,2}$.
In the cases under consideration, \genus\ refers
to the point in the Coulomb branch, where the full gauge group
$(G,G')$ is present. 
In general [$s\neq 0$; cf. eq. (2.5)], 
the supersymmetric index depends on the topology of the manifold, e.g. 
$\chi_{K3}$ and the topology of the gauge bundle, e.g. $n^{(1)},n^{(2)}$. 
As a consequence it does not change under deformations of the
hypermultiplet moduli space.
Thus we may do some change in the hypermultiplet moduli space by 
(un)Higgsing or moving in the instanton moduli space\foot{In fact, 
Higgsing and changing the gauge bundle are on the same footing.}. Both effects
result in a change of the
gauge groups $(G,G')$. This way we may very easily move from models with
standard instanton embedding ($SU(2)$ bundle in one $E_8$) 
to non--standard embeddings, if compatible with the index.
However, the net number of vector- and hypermultiplets\foot{This follows
from cancellation of the $R^4$ anomaly in six dimensions.
In six dimensions: $N_H-N_V=244$ for the models we are considering, i.e.
with one tensor multiplet $N_T=1$.} $N_H-N_V=240$
and the instanton numbers $n^{(1)},n^{(2)}$ do not change in perturbation
theory. There are restrictions on the possibility of maximal Higgsing
away $(G,G')$, which depend on the number $n$:
For $(n^{(1)},n^{(2)})=(24,0)$, i.e. $n=-12$ the second $E_8$ cannot
be broken at all, i.e. $G'=E_8$. For $n=0,1,2$ complete Higgsing is
possible. On the other hand, e.g. for $-n=3,4,6,8$, there are too few 
instantons or too less matter in the second $E_8$ to break it completely, 
thus ending with the terminal gauge groups
$G=SU(3),SO(8),E_6,E_7$, respectively. In the cases $-n=9,10,11$, 
i.e. $n^{(2)}<4$, 
the instantons on the second $E_8$ are not stable and become small,
because $D$--terms in six dimensions do not allow them
to acquire a finite size \witten. 
But then they also cannot break $E_8$, thus $G'=E_8$. 
The small instanton dynamics 
corresponds to a tensionless non--critical string theory in $d=6$ with
$E_8$ chiral algebra \ganor\kmv.

The index \genus\ is worked\foot{
Actually, in an orbifold limit of $K3$, which is smoothly connected 
to $K3$ by blowing up. This limit corresponds to going to 
special points in the hypermultiplet moduli space.} $(s=0)$ out in \hm\ for
the case of $SU(2)$--bundles in the $E_8$'s with instanton numbers
$n^{(1)},n^{(2)}$: 
\eqn\basic{
Z_{K3}(q)=-2\lf[{n^{(1)}\o 24}{E_6E_4 \o \eta^{24}}+
{n^{(2)}\o 24}{E_4E_6 \o \eta^{24}}\ri]=-2{E_4E_6 \o \eta^{24}}\ .}

The relevant objects appearing in the one--loop corrections 
$\Delta_a$ to the gauge kinetic term
$k_a g^{-2}_{\rm string}F^a_{\mu\nu}F^{a\mu\nu}$ 
in the low--energy effective action \vk
\eqn\vadim{
\Delta_a=\int_\Gamma {d^2\tau \o \tau_2}\lf[{\Bc}_a(\tau,\ov\tau)-
b_a\ri]}
for the gauge couplings $G_a=G,G'$ and the indices \basic\ are
\eqn\Kth{
\eqalign{{\Bc}_{G}(\tau,\ov\tau)&=-{1\o 12}\lf[\lf(E_2-{3\o \pi \tau_2}\ri)
{E_4E_6\o \eta^{24}}-{n^{(1)}\o 24}{E_4^3\o\eta^{24}}-
{n^{(2)}\o 24}{E_6^2\o\eta^{24}}\ri]Z_{2,2}\cr
{\Bc}_{G'}(\tau,\ov\tau)&=-{1\o 12}\lf[\lf(E_2-{3\o \pi \tau_2}\ri)
{E_4E_6\o \eta^{24}}-{n^{(1)}\o 24}{E_6^2\o\eta^{24}}-
{n^{(2)}\o 24}{E_4^3\o\eta^{24}}\ri]Z_{2,2}\ ,}}
respectively for the case $s=0$.
Due to the unique form of the $K3$ supersymmetric index 
these functions are unique for all kinds of gauge threshold corrections
without Wilson lines and lead to quite generic expressions for
the one--loop corrections \dklz\hm\kkpr.  
On the other hand, more general instanton backgrounds
give rise to a much wider class of gauge threshold corrections
than considered in \dklz.
The modular functions ${\cB}_a$ appearing
in \Kth\ are just descendents of the genus $Z_{K3}$ \genus. 
This means, that 
they are obtained from it by a $q$--derivative, which leads to the 
$(F^a)^2$--part.

In the following we want to include Wilson lines. 
Wilson lines will allow us to read off the different instanton
numbers $n^{(1)},n^{(2)}$ of a $K3$ supersymmetric index with $SU(2)$--bundles.
In the case with one Wilson line modulus $V$ w.r.t. $T^2$ ($s=1$) in an 
$SU(2)$ subgroup of the second $E_8'$ the supersymmetric index \genus\ takes 
the form \ccl
\eqn\genusB{
{\cal Z}(q,\ov q)=Z_{K3}\otimes Z_{3,2}(q,\ov q)
=-2\lf({n^{(1)}\o 24}{E_6E_{4,1} \o \eta^{24}}+{n^{(2)}\o 24}
{E_4E_{6,1} \o \eta^{24}}\ri)\otimes Z_{3,2}(q,\ov q)\ ,}
with instanton numbers $(n^{(1)},n^{(2)})=(12-n,12+n)$
w.r.t. $SU(2)$--bundles in both $E_8$ and:
\eqn\Z{\eqalign{
Z_{3,2}(q,\ov q,T,U,V)&=\sum_{m_i,n^i,k} 
e^{2\pi i\tau({1\o 4} k^2+m_1n^1+m_2n^2)}\ 
e^{-2\pi\tau_2 |p_R|^2}\cr
Y&=-{1\o 4}[(T-\ov T)(U-\ov U)-(V-\ov V)^2]=T_2U_2-V_2^2\cr
p_R&={1\o \sqrt{2Y}}[(TU-V^2)n^2+Tn^1-Um_1+m_2+kV]\ .\cr}}
In that case, the functions \Kth\ change to ($s=1$):
\eqn\KthB{\kern-1.5em
\eqalign{{\Bc}_{G}&=-{1\o 12}\lf[\lf(E_2-{3\o \pi \tau_2}\ri)
\lf({n^{(1)}\o 24}{E_{4,1}E_6\o\eta^{24}}+{n^{(2)}\o 24}
{E_4E_{6,1}\o\eta^{24}}\ri)
-{n^{(1)}\o 24}{E_4^2E_{4,1}\o\eta^{24}}-{n^{(2)}\o 24}{E_6E_{6,1}\o\eta^{24}}
\ri]\otimes Z_{3,2}\cr
{\Bc}_{G'}&=-{1\o 12}\lf[\lf(E_2-{3\o \pi \tau_2}\ri)
\lf({n^{(1)}\o 24}{E_{4,1}E_6\o\eta^{24}}+{n^{(2)}\o 24}
{E_4E_{6,1}\o\eta^{24}}\ri)
-{n^{(1)}\o 24}{E_6E_{6,1}\o\eta^{24}}-{n^{(2)}\o 24}
{E_{4,1}E_4^2\o\eta^{24}}\ri]\otimes Z_{3,2}\ .}}
The mixing of $Z_{3,2}$ with $Z_{K3}$ by the Wilson line $V$
is formally denoted by the product $\otimes$ and explained in appendix A.
The $G,G'$--beta function coefficients may be determined to:
\eqn\betasieben{\eqalign{
b_{G}^{N=2}&=12-6n\ ,\cr
b_{G'}^{N=2}&=\lf\{12+6n\ \ ,\ \ V=0 \atop 12+4n\ \ ,\ \ V\neq 0 \ri.\ .}}
Of course, the residual group $G'$ and thus $b_{G'}^{N=2}$ 
depend on the choice for the Wilson line. Statements made earlier (for
$s=0$) about
changes in the hypermultiplet moduli space remain valid.
Since we want at least one $SU(2)$ gauge group in $G'$, in the course of
Higgsing, the Wilson line modulus $V$ corresponding to the $U(1)$ Cartan
subalgebra remains a flat direction.

The $\beta$--function coefficients \betasieben\ may be
viewed  (for $V=0$) in the anomaly--polynomials $I_m$ and/or gauge 
kinetic terms in six dimensions 
(without additional tensors, which relax the factorization condition) 
\gsw\erler\sw\dmw
\eqn\sixanomalypolynom{
I_8=I_4\tilde I_4=(\tr R^2-v_G\tr F_G^2-v_{G'}\tr F_{G'}^2)
\lf(\tr R^2-v_G{n^{(1)}-12\o 12}\tr F_G^2-v_{G'}
{n^{(2)}-12\o 12}\tr F_{G'}^2\ri)\ ,}
with the Kac--Moody levels $v_a=2,1,{1\o 3},{1\o 6},{1\o 30}$ for 
$a=SU(N),SO(2N),E_6,E_7,E_8$, respectively.

The first piece of \KthB\ is universal and appears also in the gravitational
one--loop correction
\eqn\grav{
\Delta_{grav.}=-{1\o 96\pi^2}\int {d^2\tau \o \tau_2}
\lf[(-2\hat{E}_2)\lf({n^{(1)}\o 24}
{E_{4,1}E_6\o\eta^{24}}+{n^{(2)}\o 24}{E_4E_{6,1}\o\eta^{24}}\ri)
\otimes Z_{3,2}-b^{4d,N=2}_{grav.}\ri]\ ,}
with the gravitational $\beta$--function coefficient:
\eqn\betagrav{
b_{grav.}^{4d,N=2}=48+2(N_H-N_V)=\lf\{528\ \ ,\ \ V=0 \atop 
                                      468-24n\ \ ,\ \ V\neq0 \ri.\ .}
The two cases differ by $2(N_H'-N_V')=60+24n$, with:
\eqn\spectrum{
N_V'=c_n(-1)=2\ \ ,\ \ N_H'=-c_n(-1/4)=32+12n\ .}
The numbers $c_n(N)$ are the coefficients in the expansion of the
$K3$ index in \genusB\ (cf. also appendix A)
\eqn\expansion{
\hat{Z}_{K3}(q)=2\lf({n^{(1)}\o 24}
{\hat{E}_{4,1}E_6\o\eta^{24}}+{n^{(2)}\o 24}{E_4\hat{E}_{6,1}\o\eta^{24}}\ri)=
\sum_{N\in \IZ\atop N\in \IZ+{3\o 4}}     c_n(N) q^N\ .}

\newsec{Gauge threshold corrections in orbifold limits of $K3\times T^2$}

This section is devoted to a detailed analysis of the supersymmetric index
\genus\ for $K3$ orbifolds\foot{Non--freely
acting discrete twists $\IZ_\nu$ on the $T^4$--torus. 
We do not consider additional shifts in $T^2$.}.
However, since the $K3$ index may be calculated in an
orbifold limit, we also obtain results for the $K3$ index, itself.
Different orbifold limits of $K3$, correspond to different points in the $K3$
moduli space. The orbifold limits are classified by their choice of
$T^4$--twists $\IZ_\nu$ and accompanied shifts $(\gamma,\tilde\gamma)$
in the gauge degrees of freedom. 
It is one of the main results\foot{This was demonstrated for
standard--embedding orbifolds in \mm, whereas we also turn 
to non--standard--embedding} of this section to
show, that different orbifold limits of $K3$, i.e. different sets of 
$(\nu,\gamma,\tilde\gamma)$ or different points in the $K3$--moduli
space lead to the same supersymmetric index 
as long as their instanton numbers $n$ are the same.
For $K3$ vacua with $SU(2)$ bundles, it is given by \basic\ for
vanishing Wilson lines and \genusB\ for the
case with one Wilson line $V$. In particular, models of one
Higgs chain, which have the same orbifold limit, have the same genus.  
For concreteness we only consider two cases: Vanishing 
Wilson lines $V_i=0$, i.e. in this case we sit at that point in the 
vector moduli space, where the (non) Abelian gauge group is fully 
established. Then on the typeIIA side, Higgsing (with fundamental 
charged matter) corresponds to extremal transitions between topological 
distinct CY vacua \per\sonne. 
The second case with one non--vanishing Wilson line may be obtained from the 
first one by going to the Coulomb branch of an $SU(2)$ subgroup of the full
group.
The instanton numbers can be related to the shifts
$(\gamma,\tilde\gamma)$ (cf. section 3.6). 
An immediate consequence is that similar unifying statements hold for 
all kinds of physical amplitudes, which are given by the
supersymmetric index and only depend on the vector multiplets, e.g. gauge
and gravitational threshold corrections. However, threshold
corrections to a gauge group, which exist only in the orbifold limit, cannot
be similarly unified (cf. section 3.4). 
One has to be careful in the choice of non--vanishing Wilson lines and 
shifts $\gamma,\tilde\gamma$:
Only Wilson lines $V_i$ w.r.t. to the $K3$ gauge groups $G,G'$
represent directions in the smooth $K3$, which are independent on the 
hypermultiplet moduli space.
Nonvanishing Wilson lines $V_i\neq 0$ mean, that we stick to a region of the
vector multiplet moduli space, where the corresponding $SU(2)$ gauge symmetry
is in the Coulomb phase. On the other hand, when we go back to the $K3$
limit, we move in the hypermultiplet moduli space and would Higgs
away this $U(1)$, if it belonged to the $K3$--bundle. In that case,
the result does depend on the specific orbifold limit.
The non--trivial instanton background $H,H'$ of an orbifold 
may change (however not the instanton numbers)  
when blowing up to a smooth $K3$. E.g. (as we will see later) there is
the $\IZ_3$ orbifold with gauge group $U(1)^2\times SO(14)^2$ and
instanton numbers $(12,12)$. It has $U(1)$ instanton backgrounds, since the 
gauge group has rank $16$. 
On the other hand, since it does not appear in the list of \gsw, which
shows all possible Abelian backgrounds of $K3$, we conclude that the
$U(1)$--bundles convert to non--Abelian ones in the course of blowing
up the $K3$. This is also manifest in the form \genusB\ of the $K3$ 
elliptic genus (cf. table 2).
Let us also mention, that an orbifold cannot always be blown up to a
smooth $K3$ manifold. This happens, whenever there are not enough massless
oscillator modes, which serve as blowing up operators (see
e.g. \afiuv\ for examples).

The supersymmetric index for an orbifold of twist order $\nu$ and gauge
shifts $(\gamma,\tilde\gamma)$ takes for generic points in the Coulomb
moduli space $(T,U,V_i)$ the form \japan\mm:
\eqn\vadimz{
{\Zc}(q,\bar q)=-{1\o 4}\eta^{-20}(\tau)\sum\limits_{(a,b)} {1\o \nu}
Z^{(a,b)}_{K3}(q)Z^{(a,b)}_{18,2}(q,\ov q)\ ,}
with the corresponding twisted partition functions ($a,b=0,\ldots,\nu-1$):
\eqn\char{\eqalign{
Z^{(a,b)}_{K3}(q)&=k_{(a,b)}q^{-{a^2\o \nu^2}}\eta^2(\tau)
\Theta_1^{-2}\lf(\tau{a\o \nu}+{b\o \nu},\tau\ri) \cr
Z^{(a,b)}_{18,2}(q,\ov q,T,U,V_i)&=e^{-2\pi i {ab \o \nu^2}
(\gamma^2+\tilde \gamma^2)}
\sum_{p\in\Gamma^{18,2}+{a\o \nu}(\gamma+\tilde \gamma)}e^{2\pi i {b\o \nu} 
p(\gamma+\tilde \gamma)}\ q^{\h p_L^2}\ov{q}^{\h p_R^2} \ .\cr}}
We introduced:
\eqn\thetas{\eqalign{
\th\lf[a\atop b\ri](\tau)&=\sum_{k\in \IZ} q^{\h(k+\h a)^2} 
e^{i\pi(k+\h a)b}\cr
\Theta_1(z,\tau)&=i\sum_{k\in \IZ}(-1)^kq^{\h(k-\h)^2}e^{2\pi i
z(k-\h)}\ .\cr}}
At the special points $V_i=0$ the lattice sum $Z^{(a,b)}_{18,2}$ 
factorizes into 
$$Z^{(a,b)}_{18,2}=Z_{2,2}(q,\ov q)e^{-2\pi i {ab \o \nu^2}
(\gamma^2+\tilde \gamma^2)}Z^{(a,b)}_{E_8}(q)Z^{(a,b)}_{E_8'}(q)\ ,$$ 
with
\eqn\eight{
Z_{E_8}^{(a,b)}(q)=\h\sum_{\alpha,\beta}
e^{-i\pi\beta {a\o \nu}\sum\limits_{I=1}^8\gamma^I}\prod_{I=1}^8
\th\lf[{\alpha+2{a\o \nu} \gamma^I}\atop{\beta+2{b\o \nu} \gamma^I}\ri]}
and an analog expression for $Z^{(a,b)}_{E_8'}$.
For the case with one non--vanishing Wilson line
$V:=V_{16}\neq 0$, the second lattice function becomes
\eqn\eightC{
Z^{(a,b)}_{E_8'}(q,y)=\h\sum_{\alpha,\beta}e^{-i\pi\beta {a\o \nu}
\sum\limits_{I=1}^8\gamma^I}\theta\lf[\alpha \atop\beta\ri](q,y)^2
\prod_{I=1}^6\th\lf[{\alpha+2{a\o \nu} \gamma^I}\atop{\beta+2{b\o \nu} 
\gamma^I}\ri]}
with the Jacobi functions defined in appendix \appA.
For the models we are interested in, the explicit expression for the functions
${\cal B}_A$ in \vadim\  is easily derived from \eight\ and \eightC.
The charge $Q_A$ insertion is accomplished by the respective
$q$--derivative on the $\theta$--function corresponding
to the relevant $U(1)$--charge. Following \mm, we define:
\eqn\power{
-{1\o 4}{1\o\nu}e^{-2\pi i{ab\o \nu^2}(\gamma^2+\tilde
\gamma^2)}\eta^{-20}(\tau)Z^{(a,b)}_{K3}(\tau)
=:\sum_{m\geq -1}c_A^{(a,b)}(m) q^m\ .}
With these definitions 
the $\beta$--function coefficients \betasieben,
which ensure that \vadim\ remains IR--finite, are determined to be: 
\eqn\betas{
b^{N=2}_A\equiv\lim_{\tau_2\ra\infty}{\Bc}_A(\tau,\ov\tau)=
\sum_{n_i}\sum_{(a,b)}  e^{2\pi i{b\o \nu} 
[n+{a\o \nu}(\gamma+\tilde\gamma)](\gamma+\tilde\gamma)}
{(n_iQ_i)^2 \o Q^2_A} c_A^{(a,b)}\lf(-{1\o 2} n_i^2\ri)\ .}

\subsec{Automorphic forms and gauge threshold corrections}

The threshold corrections $\Delta_a$ can be split into three pieces \nsz:
\eqn\three{
\Delta_a=b_a\triangle-G^{(1)}+\sigma\ .}
The first term depends on the gauge group under consideration and
it is entirely given in terms of $SO(2+s,2,\IZ)$ modular functions.
It is that piece\foot{In the following we call $\triangle$ automorphic form,
although it is $\triangle+\ln(Y\kappa)$, which constitutes an automorphic
form.}, which gives rise to automorphic forms.
A prominent example is the Dedekind $\eta$--function for the
case $s=0$. With the corresponding logarithmic singularity arising from the
Kaluza--Klein states becoming massless at $T\ra i\infty$ it reads\foot{
Essential modifications arise in the case of $SL(2,\IZ)$ subgroups \mse.}
\dklz:
\eqn\dixon{
\triangle=-\ln[\kappa(-iT+i\ov T)(-iU+i\ov U)]|\eta(T)|^4|\eta(U)|^4\ ,}
with the regularization constant 
$\kappa={8\pi \o 3\sqrt 3}e^{1-\gamma_E}$ and $\gamma_E$ being the
Euler--Mascheroni constant.
The correction $G^{(1)}$ is the one--loop correction to the K{\"a}hler 
potential \dfkz. 
Because of supersymmetry it also appears in the integral \vadim.
Finally, $\sigma$ summarizes additional moduli--dependent corrections.
They are the subject of \kk\ for the case without Wilson lines 
and of \nsz\ when including Wilson lines.
Therefore, to isolate the automorphic form $\triangle$ in \three, we focus 
(in this subsection) on
the difference of two distinct gauge groups. The
integral \vadim\ can be formally evaluated \mm\ and gives for the
difference of two gauge groups $G_A$ and $G_{A'}$
\eqn\product{
\triangle:={{\Delta_A-\Delta_{A'}}\o{b^{N=2}_A-
b^{N=2}_{A'}}}=-\ln(Y\kappa) -
4\re\lf\{\sum_{\al>0}\ \ln e^{\pi i \rho y}\lf(1-e^{2\pi i \al
y}\ri)^{{d_A(\al)-d_{A'}(\al)\o b^{N=2}_A-
b^{N=2}_{A'}}}\ri\}\ ,}
with
\eqn\ds{
d_A(\al)={(n_iQ_i)^2 \o Q^2_i}\sum_{a,b}e^{2\pi i {b \o \nu} 
[n+{a\o \nu}(\gamma+\tilde \gamma)]
(\gamma+\tilde \gamma)}c_A^{(a,b)}\lf(kl-\h n_i^2\ri)\ ,}
and $\alpha=(k,l,n_i)$, $Y=-{1\o 4}[(T-\ov T)(U-\ov U)-\sum\limits_{i=1}^{16}
(V_i-\ov V_i)^2]$ and $\alpha y=kT+lU+n_iV_i$. The gauge group
dependent numbers $\tilde d_A$ appearing in the vector 
$\rho={1\o 4(b^{N=2}_A-b^{N=2}_{A'})}(\tilde d_A-\tilde d_{A'})$
may be looked in \mm. The sum $\alpha>0$ runs over all positive
lattice vectors
$(i)\ k>0,l\in \IZ,n_i\in \IZ$,\ $(ii)\ k=0,l>0,n_i\in \IZ$,
\ $(iii)\ k,l=0,n_i>0$.
The expression \product\ seems to depend on the orbifold twist $\nu$,
the underlying gauge embeddings $(\gamma,\tilde \gamma)$ and finally on the 
two gauge groups 
between which the difference is taken.
In the following we want to demonstrate that this is an artifact.
In fact, we will see (for the case $s=0,1$), that the r.h.s. of \ds\ 
gives rise to one generic automorphic form (or certain linear combinations) 
of $SO(2+s,s,\IZ)$, being independent of the orbifold details.

Let us pursue this idea further.
For concreteness we will specialize to the one Wilson line $V_{16}:=V$ case.
By looking at the perturbative 
duality symmetry $SO(3,2,\IZ)$ and at the singularity structure in the
moduli space of \vadim, the gauge group--dependent part
$b_a^{N=2}\triangle$ of  
threshold corrections $\Delta_a$ involving one Wilson line modulus 
could be derived in \msf.
Two cases of physical gauge couplings are relevant \msf.
In the first case, no (under the considered gauge group) charged 
particles become massless for $V\ra 0$ and
the form of these thresholds is given by\foot{See e.g. the appendix of 
\nsz\ and \msf\ for further information about Siegel forms. The relevance of
Siegel modular forms in the context of string one--loop corrections
was first observed in \msf.}:
\eqn\dz{
\triangle=-{1\o 12}\ln(\kappa Y)^{12}|\chi_{12}|^2\ .}
In the second case, some particles,
charged under the gauge group under consideration, become massless
for $V\ra 0$. This means that the effective one--loop correction 
develops a logarithmic singularity in this limit, since those particles, 
which run in the loop and become massless, have been integrated out.
The form of these thresholds is given by
\eqn\de{
\triangle=-{1\o 10}\ln(\kappa Y)^{10}|\chi_{10}|^2\ .}
Not any universal contribution are included in these functions.
Both thresholds are entirely due to the gauge group dependent 
part of the charge insertion $Q_a$ in \vadim.
I.e. they may be determined by considering a difference $\triangle$ of two 
gauge groups thresholds $\Delta_{G_a},\Delta_{G_a'}$. 
In the first case, a difference involving two 
gauge groups, which are not enhanced at special points in the moduli
space. In the second case, two gauge groups, which are both enhanced
at $V\ra 0$. 
The appearance of the $SO(3,2,\IZ)$ automorphic form $\chi_{12}$
in \dz\ is plausible as it gives the correct result \dixon\ in the limit
$V\ra 0$, due to $\lf.\chi_{12}\ri|_{V=0}=\eta^{24}(T)\eta^{24}(U)$. 
Moreover, the expression $\Delta_A-\Delta_{A'}$ can be expanded in
powers of $V$  (cf. appendix C)
\eqn\productz{\hskip-0.5cm
\eqalign{\triangle\equiv
{{\Delta_A-\Delta_{A'}}\o{b^{N=2}_A-b^{N=2}_{A'}}}&=
-\ln (\kappa Y)-{1\o 12}|\eta^{24}(T)\eta^{24}(U)|^2 |1+12 V^2 \p_T 
\ln\eta^2(T) \p_U\ln\eta^2(U)|^2\ +\ldots\ \cr
&=-{{1}\o{12}}\ln (\kappa Y)^{12} |\chi_{12}|^2\ ,\cr}}
which agrees with the lowest order of \dz.
The form of \de\ ensures the correct logarithmic behaviour at $V\ra 0$
($\chi_{10}\ra V^2\eta^{24}(T)\eta^{24}(U)$).
Finally, in \product\ one may also consider the case of two
gauge groups, where one gets enhanced for $V\ra 0$ and the other does not:
\eqn\dez{
\triangle=-{1\o 12}\ln(\kappa Y)^{12}|\chi_{12}|^2
+\lf[1-{b^{N=2}(V=0)\o b^{N=2}(V\neq 0)}\ri]
\ln\lf|{\chi_{10}^{1/2} \o \chi_{12}^{5/12}}\ri|^2\ .}
This represents a case, where automorphic forms show up in a linear 
combination,
with a gauge group dependent factor. This dependence may be eliminated 
if one considers a second pair of gauge groups.

In the next subsection we will see that differences of 
gauge threshold corrections \product\ of any $K3$ orbifold  always take
the form \de\ or \dez. This result is quite intriguing, since
\product\ depends on the specific gauge embedding
$(\gamma,\tilde\gamma)$ of
the orbifold and its spectrum, whereas \de\ and \dez\ are
model--independent. All model dependence  goes into the
$\beta$--function coefficients, which in \three\ appear just as prefactors.
We are not able to reproduce \dz\ from \product.
We would have to choose a pair of two non--singular couplings.
However, looking at (3.18), such a pair only exists
for the cases $(iii)$ and $(v)$. They have the same $\beta$--function
coefficients, a fact which formula \product\ does not allow.
We believe, that this failure is no accident, since it would lead to a
product formula for $\chi_{12}$, whereas its divisors do not have any 
simple form \borch. 

\subsec{Standard orbifold limits of $K3\times T^2$}

In this section we work out 
\vadim,\genusB\ and in particular \product\ for standard and non--standard
orbifolds $K3$ orbifold limits.
Since the result \dixon\ holds for all types of orbifolds
of $K_3\times T^2$, i.e. in the case of both standard  
and non--standard embedding of the twist
into the gauge degrees of freedom, we expect
this to hold also for the case when one non--trivial gauge background
field, i.e. one Wilson line $V$ is turned on. 
It is one of the aim of this section 
to obtain general expressions for $\Delta$ in those cases.

In the following we consider the standard--embeddings
\eqn\stand{
\gamma^I=(1,-1,0,0,0,0,0,0)\ \ ,\ \ {\tilde \gamma}^I=(0,0,0,0,0,0,0,0)
\ \ ,\ \ \nu=2,3,4,6,}
with the N=2 gauge groups $SU(2)\times E_7\times E_8$ for $\nu=2$\ 
($b^{N=2}_{SU(2)}=b_{E_7}^{N=2}=84,\ b_{E_8}^{N=2}=-60$) and
$U(1)\times E_7\times E_8$ for the others.
In these cases eq. \vadimz\ simplifies\foot{For all N=2 orbifolds 
(see the following tables) we expanded \vadimz\ in a power series 
in $q$ and $y$ and found agreement with \genusB\ up to an arbitrary
high order.} drastically.
In fact, it reduces
to the form \genusB\ with $(n^{(1)},n^{(2)})=(24,0)$.
Moreover the threshold corrections take the form \Kth\ and \KthB\ with
the same instanton numbers. 

\subsec{Non--standard orbifold limits of $K3\times T^2$}

For concreteness, let us discuss seven cases of non--standard embeddings.

{\vbox{\ninepoint{
\def\ss#1{{\scriptstyle{#1}}}
$$
\vbox{\offinterlineskip\tabskip=0pt
\halign{\strut\vrule#
&~$#$~\hfil
&\vrule#
&~$#$~\hfil
&~$#$~\hfil
&~$#$~\hfil
&\vrule#
&~$#$~\hfil
&\vrule#
\cr
\noalign{\hrule}
&
\ 
&&
\ss {\nu}
&
\ss{\gamma}
&
\ss{\tilde\gamma}
&&
\ss{\rm \ perturbative\ gauge\ group}
&
\cr
\noalign{\hrule}
&
\ss i
&&
\ss 2
&
\ss{ (1,-1,0,0,0,0,0,0)}
&
\ss{ (2,0,0,0,0,0,0,0)}
&&
\ss{ SU(2)\times E_7\times SO(16)'}
&
\cr
&
\ss{ ii}
&&
\ss{ 3}
&
\ss{ (1,-1,0,0,0,0,0,0)}
&
\ss{ (2,1,1,0,0,0,0,0)}
&&
\ss{ U(1)\times E_7\times SU(3)'\times E_6'}
&
\cr
&
\ss{ iii}
&&
\ss{ 3}
&
\ss{ (2,0,0,0,0,0,0,0)}
&
\ss{ (2,0,0,0,0,0,0,0)}
&&
\ss{ U(1)\times SO(14)\times U(1)'\times SO(14)'}
&
\cr
&
\ss{ iv}
&&
\ss{ 3}
&
\ss{ (2,1,1,0,0,0,0,0)}
&
\ss{ (2,1,1,1,1,0,0,0)}
&&
\ss{ SU(3)\times E_6\times SU(9)'}
&
\cr
&
\ss{ v}
&&
\ss{ 4}
&
\ss{ (2,2,2,0,0,0,0,0)}
&
\ss{ (3,1,1,1,1,1,0,0)}
&&
\ss{ SU(4)\times  SO(10)\times  SU(2)'\times SU(8)'}
&
\cr
&
\ss{ vi}
&&
\ss{ 4}
&
\ss{ (1,1,1,-3,0,0,0,0)}
&
\ss{ (1,1,-2,0,0,0,0,0)}
&&
\ss{ SU(4)\times SO(10)\times U(1)'\times SU(2)'\times E_6'}
&
\cr
&
\ss{ vii}
&&
\ss{ 6}
&
\ss{ (1,1,1,1,-4,0,0,0)}
&
\ss{ (1,1,1,1,1,-5,0,0)}
&&
\ss{ U(1)\times  SU(4)\times SU(5)\times SU(2)'\times SU(3)'\times SU(6)'}
&
\cr
\noalign{\hrule}}
\hrule}$$
\vskip-10pt
\noindent{\bf Table 1:}
{\sl Examples of non--standard orbifold limits of heterotic $K3\times T^2$
compactifications $(\nu,\gamma,\tilde\gamma)$ and 
their perturbative gauge group.}
\vskip10pt}}}
Models $(ii),(vi),(vii)$ correspond to the $r=10,8,4$ chains, respectively 
discussed in \afiq. Actually in total, there are $2$ different
embeddings for $\IZ_2$, $5$ for $\IZ_3$, $12$ for $\IZ_4$ (table 3)
and $61$ for $\IZ_6$ (table 8).

With the relation
\eqn\Ntwobeta{
b_a^{N=2}=2\Tr_H(Q^2_a)-2\Tr_V(Q_a^2)\ ,}
where $Q_a$ is any generator\foot{In the case of non--vanishing Wilson line
$V\neq 0$, generators $Q_a$, which do not survive the Wilson line projection, 
are excluded. In that case, the $\beta$--function coefficient refers to 
the surviving (smaller) gauge group.}  of the group $G_a$, 
we determine the following N=2 $\beta$--function coefficients
\eqn\Betas{\eqalign{
  (i)\  &\  b^{N=2}_{E_7}=-12\ \ ,\ \ 
b_{SO(16)'}^{N=2}=\lf\{36\ ,\ V=0\atop 28\ ,\  V\neq 0\ri.\cr 
  (ii)\ &\  b^{N=2}_{E_7}=-24\ \ ,\ \ 
b_{E_6'}^{N=2}=\lf\{48\ ,\ V=0\atop 36\ ,\ V\neq 0\ri.\cr
  (iii)\ &\ b^{N=2}_{SO(14)}=12\ \ ,\ \ b_{SO(14)'}^{N=2}= 12\cr
  (v)\ &\ b^{N=2}_{SO(10)}=12\ \ ,\ \ b_{SU(8)'}^{N=2}=12\cr
  (vi)\  &\ b^{N=2}_{SO(10)}=36\ \ ,\ \ b_{E_6'}^{N=2}=
           \lf\{-12,\ V=0\atop -4\ ,\ V\neq 0\ri.\cr
  (vii)\  &\ b^{N=2}_{SU(5)}=24\ \ ,\ \ b_{SU(6)'}^{N=2}=
           \lf\{0\ ,\ V=0\atop 4\ ,\ V\neq 0\ ,\ri.\cr}}
respectively.
Our results can be easily converted to other embeddings.
Interestingly, an explicit calculation shows that in all seven cases 
the supersymmetric index \vadimz\ reduces to the  expression
\genusB\ with instanton numbers $(n^{(1)},n^{(2)})$ referring to the
$SU(2)$ bundles in $E_8\times E_8$. This is also the case for the
other orbifold limits of $K3$ (cf. e.g. table 3 for all $T^4/\IZ_4$
and appendix D for all $T^4/\IZ_6$ orbifolds).

\vskip0.5cm
{\vbox{\ninepoint{
\def\ss#1{{\scriptstyle{#1}}}
$$
\vbox{\offinterlineskip\tabskip=0pt
\halign{\strut\vrule#
&~$#$~\hfil
&\vrule#
&~$#$~\hfil
&~$#$~\hfil
&~$#$~\hfil
&\vrule#
\cr
\noalign{\hrule}
&
\ 
&&
\ss{ (n^{(1)},n^{(2)})}
&&
\ss{ n}
&
\cr
\noalign{\hrule}
&
\ss{ i}
&&
\ss{ (8,16)}
&&
\ss{ 4}
&
\cr
&
\ss{ ii}
&&
\ss{ (6,18)}
&&
\ss{ 6}
&
\cr
&
\ss{ iii}
&&
\ss{ (12,12)}
&&
\ss{ 0}
&
\cr
&
\ss{ iv}
&&
\ss{ (9,15)}
&&
\ss{ 3}
&
\cr
&
\ss{ v}
&&
\ss{ (12,12)}
&&
\ss{ 0}
&
\cr
&
\ss{ vi}
&&
\ss{ (16,8)}
&&
\ss{ -4}
&
\cr
&
\ss{ vii}
&&
\ss{ (14,10)}
&&
\ss{ -2}
&
\cr
\noalign{\hrule}}
\hrule}$$
\vskip-10pt
\centerline{\noindent{\bf Table 2:}
{\sl The instanton numbers $(n^{(1)},n^{(2)})$ of the 
previous examples.}}
\vskip10pt}}}

Of course, this is quite remarkable as in \vadimz\
we are summing over all various twisted sectors. Altogether this results
in \genusB. On the other hand, this fact may be understood 
from the point of view of modular functions: The expression \vadimz\ 
is given by a modular function of weight $-2$ and a certain pole structure
dictated by the states becoming massless for $\tau\rightarrow -i\infty$ 
(tachyon)
and in the IR $\tau\rightarrow i\infty$. This fixes the form of \vadimz. 
Therefore we conclude, that \genusB\ and \KthB\ are the most general 
expressions for the index and gauge threshold corrections in orbifold
limits of $K3$, respectively.

{\vbox{\ninepoint{
\def\ss#1{{\scriptstyle{#1}}}
$$
\vbox{\offinterlineskip\tabskip=0pt
\halign{\strut\vrule#
&~$#$~\hfil
&~$#$~\hfil
&~$#$~\hfil
&\vrule#
&~$#$~\hfil
&\vrule#
&~$#$~\hfil
&\vrule#
&~$#$~\hfil
&\vrule#
\cr
\noalign{\hrule}
&
\ss {\gamma}
&&
\ss {\tilde\gamma}
&&
\ss {\rm \ perturbative\ gauge\ group}
&&
\ss{ (n^{(1)},n^{(2)})}
&
\cr
\noalign{\hrule}
&
\ss{ (1,1,0,0,0,0,0,0)}
&&
\ss{ (0,0,0,0,0,0,0,0)}
&&
\ss{ U(1)\times E_7\times E_8'}
&&
\ss{ (24,0)}
&
\cr
&
\ss{ (1,1,0,0,0,0,0,0)}
&&
\ss{ (2,2,0,0,0,0,0,0)}
&&
\ss{ U(1)\times E_7\times SU(2)' \times E_7'}
&&
\ss{ (12,12)}
&
\cr
&
\ss{ (1,1,0,0,0,0,0,0)}
&&
\ss{ (4,0,0,0,0,0,0,0)}
&&
\ss{ U(1)\times E_7\times SO(16)'}
&&
\ss{ (16,8)}
&
\cr
&
\ss{ (1,1,0,0,0,0,0,0)}
&&
\ss{ (1,1,1,1,1,1,1,-1)}
&&
\ss{ U(1)\times E_7\times U(1)'\times SU(8)'}
&&
\ss{ (6,18)}
&
\cr
&
\ss{ (2,1,1,0,0,0,0,0)}
&&
\ss{ (2,0,0,0,0,0,0,0)}
&&
\ss{ U(1)\times SU(2)\times E_6\times U(1)'\times SO(14)'}
&&
\ss{ (12,12)}
&
\cr
&
\ss{ (2,1,1,0,0,0,0,0)}
&&
\ss{ (2,2,2,0,0,0,0,0)}
&&
\ss{U(1)\times SU(2)\times E_6\times SU(4)' \times SO(10)'}
&&
\ss{(8,16)}
&
\cr
&
\ss{ (3,1,0,0,0,0,0,0)}
&&
\ss{ (0,0,0,0,0,0,0,0)}
&&
\ss{ U(1)\times SU(2)\times SO(12)\times E_8'}
&&
\ss{ (24,0)}
&
\cr
&
\ss{ (3,1,0,0,0,0,0,0)}
&&
\ss{ (2,2,0,0,0,0,0,0)}
&&
\ss{ U(1)\times SU(2)\times SO(12)\times SU(2)' \times E_7'}
&&
\ss{ (20,4)}
&
\cr
&
\ss{ (3,1,0,0,0,0,0,0)}
&&
\ss{ (4,0,0,0,0,0,0,0)}
&&
\ss{ U(1)\times SU(2)\times SO(12)\times SO(16)'}
&&
\ss{ (16,8)}
&
\cr
&
\ss{ (3,1,1,1,1,1,0,0)}
&&
\ss{ (2,2,2,0,0,0,0,0)}
&&
\ss{ SU(2)\times SU(8)\times SU(4)'\times SO(10)'}
&&
\ss{ (12,12)}
&
\cr
&
\ss{ (3,1,1,1,1,1,0,0)}
&&
\ss{ (2,0,0,0,0,0,0,0)}
&&
\ss{ SU(2)\times SU(8)\times U(1)'\times SO(14)'}
&&
\ss{(12,12)}
&
\cr
&
\ss{ (1,1,1,1,1,1,1,-1)}
&&
\ss{ (3,1,0,0,0,0,0,0)}
&&
\ss{ U(1)\times SU(8)\times U(1)'\times SU(2)'\times SO(12)'}
&&
\ss{ (14,10)}
&
\cr
\noalign{\hrule}}
\hrule}$$
\vskip-10pt
\noindent{\bf Table 3:}
{\sl All twelve $T^4/\IZ_4$--orbifolds with gauge twist 
$(\gamma,\tilde\gamma)$:
Their perturbative gauge group and instanton numbers
$(n^{(1)},n^{(2)})$ w.r.t. $SU(2)$--bundles.}
\vskip10pt}}}

As we have seen, the function ${\cal B}_a$ in \vadim\ is given by
\KthB\ with the topological numbers, given in the tables 2,3 and 8, 
respectively.
The $\tau$--integral in \vadim\ can be done guided by \ke\kz\ccl.
For the difference $\triangle_G-\triangle_{G'}$ of gauge threshold 
corrections, involving a non--singular gauge group $G$ and a singular
one $G'$, we obtain the proposed form \de:
In that case, 
\eqn\inthat{
{\cB}_G-{\cB}_{G'}=-{n\o 2} {\cal Z}_{K3}(q,y)\otimes Z_{3,2}\ ,}
with the $K3$ elliptic genus ${\cal Z}_{K3}(q,y)$ introduced  
eq. (A.5). Thus \ke:
\eqn\phone{
\triangle_G-\triangle_{G'}=n\ln(\kappa Y)^{10}|\chi_{10}|^2\ .}
The other expression \dez\ appears in the next subsection.

\subsec{Enhanced gauge group threshold corrections}

Since in the case of $K3$ compactifications the manifold has no 
isometries, the gauge group $(G,G')$ derives from $E_8\times E_8$ (and 
the bundle structure), only. 
However, at special points in the moduli
space, like e.g. the orbifold points of $K3$, additional gauge group
factors $(H,H')$ appear. 
The maximal\foot{This requires also a choice of 
special points in the vector multiplet moduli space $(T,U)$.} possible gauge 
group is $E_8\times E_7\times SU(2)^5$ \sch.
One feature (cf. previous sections) for threshold corrections w.r.t. $(G,G')$ 
is that, they may be equal for different choices of $(\nu,\gamma,\tilde\gamma)$
as long as they have the same instanton numbers $(n^{(1)},n^{(2)})$. 
On the other hand,
this statement no longer holds for the enhanced gauge group $(H,H')$
thresholds, which in general depend on the chosen orbifold limit.
Which gauge group arises at the orbifold limit, depends on the twist
embedding. Therefore, threshold corrections w.r.t.
those gauge groups depend on the specific form of the shift vector   
$(\gamma,\tilde\gamma)$.
As a consequence they are not expressible in terms of
$(n^{(1)},n^{(2)})$ alone.
Eq. \vadim\ then takes the form
\eqn\extra{
{\Bc}_{H,H'}=-{1\o 12}\lf[\lf(E_2-{3\o \pi \tau_2}\ri)
\lf({n^{(1)}\o 24}{E_{4,1}E_6\o\eta^{24}}+{n^{(2)}\o 24}
{E_4E_{6,1}\o\eta^{24}}\ri)
-{d_1\o 24}{E_4^2E_{4,1}\o\eta^{24}}-{d_2\o 24}{E_6E_{6,1}\o\eta^{24}}\ri]
\otimes Z_{3,2}\ ,}
with the additional coefficients $(d_1,d_2)$. 
For the examples of table 1 we determined these coefficients:

{\vbox{\ninepoint{
\def\ss#1{{\scriptstyle{#1}}}
$$
\vbox{\offinterlineskip\tabskip=0pt
\halign{\strut\vrule#
&~$#$~\hfil
&\vrule#
&~$#$~\hfil
&~$#$~\hfil
&~$#$~\hfil
&\vrule#
&~$#$~\hfil
&\vrule#
\cr
\noalign{\hrule}
&
\ 
&&
\ss{ H,H'}
&&
\ss{ (d_1,d_2)}
&&
\ss{ d}
&
\cr
\noalign{\hrule}
&
\ss{ i}
&&
\ss{ SU(2)}
&&
\ss{ (40,-16)}
&&
\ss{ -28}
&
\cr
&
\ss{ ii}
&&
\ss{ U(1)}
&&
\ss{ (30,-6)}
&&
\ss{ -18}
&
\cr
&
\ 
&&
\ss{ SU(3)'}
&&
\ss{ (18,6)}
&&
\ss{ -6}
&
\cr
&
\ss{ iii}
&&
\ss{ U(1)}
&&
\ss{ (28,-4)}
&&
\ss{ -16}
&
\cr
&
\ 
&&
\ss{ U(1)'}
&&
\ss{ (28,-4)}
&&
\ss{ -16}
&
\cr
&
\ss{ v}
&&
\ss{ SU(4)}
&&
\ss{ (20,4)}
&&
\ss{ -8}
&
\cr
&
\ss{ vi}
&&
\ss{ SU(4)}
&&
\ss{ (16,8)}
&&
\ss{ -4}
&
\cr
&
\ 
&&
\ss{ SU(2)'}
&&
\ss{ (28,-4)}
&&
\ss{ -16}
&
\cr
&
\ 
&&
\ss{ U(1)'}
&&
\ss{ (20,4)}
&&
\ss{ -8}
&
\cr
\noalign{\hrule}}
\hrule}$$
\vskip-10pt
\centerline{\noindent{\bf Table 4:}
{\sl Details $(d_1,d_2)$ for enhanced gauge group $H,H'$ 
threshold corrections.}}
\vskip10pt}}}
Interestinlgy, for all cases $d_1+d_2=24$ and
it is convenient to introduce $d$ with $d_1=12-d,\ d_2=12+d$.
For the $\beta$-functions we obtain:
\eqn\betaHH{
b_{H,H'}^{N=2}=\lf\{12-6d\ \ ,\ \ V=0 \atop 12-n-5d\ \ ,\ \ V\neq 0 \ri.\ .}
For the examples of table 4:
\eqn\BetasH{\eqalign{
  (i)\  &\  b_{SU(2)}^{N=2}=\lf\{180\ ,\ V=0\atop 148\ , \ V\neq 0\ri.\ \ ,\cr 
  (ii)\ &\  b_{U(1)}^{N=2}=\lf\{120\ ,\ V=0\atop 96\ ,\ V\neq 0\ri.\ \ ,\ \ 
            b_{SU(3)'}^{N=2}=\lf\{48\ ,\ V=0\atop 36,\ V\neq 0\ri.\cr
  (iii)\ &\ b_{U(1)}^{N=2}=\lf\{108\ ,\ V=0\atop 92\ ,\ V\neq 0\ri.\ \ ,\ \ 
            b_{U(1)'}^{N=2}=\lf\{108\ ,\ V=0\atop 92\,\ V\neq 0\ri.\cr
  (v)\ &\  b_{SU(4)}^{N=2}=\lf\{60\ ,\ V=0\atop 52\ ,\ V\neq 0\ri.\ \ ,\cr
  (vi)\  &\ b^{N=2}_{SU(4)}=36\ ,\ 
b_{U(1)'}^{N=2}=\lf\{60\ ,\ V=0\atop 56\ ,\ V\neq 0\ri.\ ,\ 
b_{SU(2)'}^{N=2}=\lf\{108\ ,\ V=0\atop 96\ ,\ V\neq 0\ .\ri.\cr}}

\subsec{Differential equation for the N=2 prepotential}

For the cases discussed in \hm\ it was shown that the one--loop
correction  $h_n$ to the prepotential of the underlying  N=2 
theory fulfils a second order differential equation.
Also in the cases at hand we can derive a differential equation
for the one loop correction $h_n$ to the N=2 prepotential $H_n$
\eqn\prep
{H_n(S,T,U,V)=S(TU-V^2)+h_n(T,U,V)+O(e^{-2\pi i S})\ ,}
with \kz
\eqn\oneprep
{h_n(T,U,V)=-{i\o{2\pi}}d_n(T,U,V)-{{1}\o{(2\pi)^4}}
\sum_{(k,l,b)>0}c_n\lf(kl-{1\o 4}b^2\ri)\Li_3(e^{2\pi i (kT+lU+bV})+const.}
and
\eqn\cubic
{d_n(T,U,V)={{1}\o{3}} U^3+({4\o 3}+n)V^3-(1+{n\o 2}) UV^2-{n\o 2}TV^2\ .}
The coefficients $c_n$ refer to the $K3$ genus $\hat{Z}_{K3}(q)$
given in \genusB\ and its coefficients \expansion.
There are ambiguities for the cubic polynomial \cubic\ due to the
fact that the holomorphic prepotential is only fixed up to quadratic pieces
in the homogeneous coordinates $\hat X^I$.
These quadratic pieces include e.g. cubics in $V$. On the other hand, 
this ambiguity can be fixed when comparing the prepotential
with the corresponding one of the typeII theory, which leads to
the form \cubic \ccl.
For the differential equation we find
\eqn\dgl
{\re\lf\{{{16\pi^2}\o{5}}(\p_T\p_U-{{1}\o{4}}\p^2_V)h_n\ri\}-
G_{N=2}^{(1)}=\Delta^{(n,d)}_\alpha+{{16}\o{5}}\re
\lf[\ln\Psi_{n,d}(T,U,V)\ri]+
b^{N=2}_{\alpha,n,d} \ln Y\ ,}
with
\eqn\univ{
G_{N=2}^{(1)}={8\pi^2\o Y}\re\lf\{h_n-\sum_{y=T,U,V} \im(y){\p\o \p y} \
h_n\ri\}\ .}
The above differential equation holds for general gauge threshold 
corrections $\Delta^{(n,d)}_\alpha$ \vadim\ which are given by
\extra\ for some $n,d$. Of course, $d=\pm n$ leads to \KthB, whereas
$d\neq \pm n$ describes the cases in table 4. 
The holomorphic function $\Psi_{n,d}$ is $(\triangle_{10}=-4\chi_{10},\   
\triangle_{35}=4i\chi_{35}$):
\eqn\psi{
\Psi_{n,d}(T,U,V)={\triangle_{35}^{1/2}\o\triangle_{10}^{1 + 1/16
n+5/16 d}}\ .}
From that differential equation and \three\ one immediately obtains\foot{
The case $n=12,d=\mp 12$ was discussed in \nsz.} an expression for
$\si$:
\eqn\sigm{
\sigma_n(T,U,V)=\re\lf\{{16\pi^2\o 5}(\p_T\p_U-{{1}\o{4}}\p^2_V)h_n-
{4\o 5\pi}\ln{\triangle_{35}\o \triangle_{10}^{3/2n+4}}
\chi_{12}^{5/4n-5/2}\ri\}\ .}


All our investigations concern the so--called $A$--models \cf. These are 
heterotic $K3\times T^2$ compactifications where the K3 of the typeIIA 
dual CYM (which is a K3 fibration) is itself a fibration with 
$E_8$--torus fiber.
In particular this means, that on the K{\"a}hler modulus 
$T$ of the heterotic torus $T^2$, 
to be identified with the K{\"a}hler modulus of the torus
fiber, the full $SL(2,\IZ)_T$ (to be embedded into $SO(2,2+s,\IZ)$
is realized \klm\byau.
The $B,C$--models have fiber tori $E_7,E_6$, respectively.
In these cases, the heterotic duality group  
of the $T$--modulus is only a subgroup of $SL(2,\IZ)_T$ and e.g. for $s=1$
the results of the previous sections involve $Sp(4,\IZ)$ subgroups \NG\car.

After Higgsing completely, the models with $n=0,1,2$ become the so--called
$STU$--models or one step before --with an $SU(2)$ gauge group in the Coulomb 
phase-- the $STUV$--models.
Their duality to typeIIA CYM, which are elliptic fibrations over the 
Hirzebruch surfaces $\IF_n$ with
\eqn\Euler{
c_n(0)=\chi(X_{\IF_n})=\lf\{ {-480\ \ ,\ \ V=0 \atop -420+24n\ \ ,\ \
V\neq 0}\ri.\ ,}
and $h_{(1,1)}=N_V-1=4$ and $h_{(2,1)}=N_H-1=214-12n$, has been checked in 
\ccl. 
The models with $n\neq 0,1,2$ have some terminal gauge groups after Higgsing 
completely (cf. also section 2) and again are related to CYM being 
elliptically 
fibered over $\IF_n$. The CY prepotential 
and the gravitational coupling, specialized to the sublocus of their 
K\"ahler moduli space, where this gauge group is fully established 
(or up to an $SU(2)$ factor, which is in the Coulomb phase
in the case of one non--vanishing Wilson line), can be linked to our 
results \oneprep. 
Dictated by the Jacobi functions, at such a point the instanton 
expansion of the CY prepotential will arrange w.r.t. to $SU(2)$ 
representations like \genusB.
 
\subsec{Geometric interpretation}

In the previous sections we have seen, that the instanton numbers 
$n^{(1)},n^{(2)}$
and the relation $n^{(1)}+n^{(2)}=24$ are important ingredients 
entering the threshold results. See e.g. \KthB\ and \extra.
On the other hand, all the information about an orbifold, e.g. the massless
spectrum and the instanton numbers are encoded
in $\nu$ and the shifts $\gamma^I,\tilde\gamma^I$. Therefore in this section
we want to express $n^{(1)},n^{(2)}$ in terms of 
$\nu,\gamma^I,\tilde\gamma^I$.
We know of two ways, to find this relation. The first one uses the 
supersymmetric index \genus\ (cf. the previous sections and tables 2,3
and 8). 
I.e. we write the index \vadimz\ in such a way \genusB, that we are able to 
read off the instanton numbers. 
This fixes the instanton numbers $n^{(1)},n^{(2)}$ completely. 
The second way uses results about small instantons at $\IZ_\nu$
orbifold singularities \intri. The
individual instanton numbers w.r.t. $E_8\times E_8'$ at an orbifold
fixed point $f_\alpha$ with gauge twists 
$\gamma_\alpha,{\tilde\gamma}_\alpha$ of 
twist order $\nu_\alpha$ are given by 
\intri\afiuv:
\eqn\index{\eqalign{
n^{(1)}_\alpha&=k_\alpha^{(1)}+\sum_{j=0}^{\nu_\alpha-1}{j(\nu_\alpha-j)\o 
4\nu_\alpha}w_j=
k^{(1)}_\alpha +\h{1\o \nu_\alpha}\sum_{I=1}^8\gamma_\alpha^I(\nu_\alpha-
\gamma_\alpha^I)\ ,  \cr
n^{(2)}_\alpha&= k_\alpha^{(2)}+\sum_{j=0}^{\nu_\alpha-1}{j(\nu_\alpha-j)\o 
4\nu_\alpha}\tilde w_j=k^{(2)}_\alpha+\h{1\o \nu_\alpha}\sum_{I=1}^8
\tilde\gamma_\alpha^I(\nu_\alpha-\tilde \gamma_\alpha^I)\ .\cr}}
Here $k_\alpha^{(1)},k_\alpha^{(2)}$ are arbitrary integers, which 
are already present for instantons in flat Euclidian $\IR^4$ and
$w_\mu$ is the number of orbifold twist eigenvalues 
$exp(2\pi i \mu/\nu_\alpha)$ in $\gamma_\alpha$. 
To finally obtain $n^{(1)},n^{(2)}$ one has to sum over all possible 
fixed point
of order $\nu_\alpha=0,\ldots,\nu-1$. Of course, $n^{(1)}+n^{(2)}=24$, which 
relates the small instanton physics at the orbifold fixed point to the 
$K3$--geometry. At one fixed point $f_\alpha$ a similar equation
holds, which relates $n^{(1)}_\alpha+n^{(2)}_\alpha$ to the Euler
number of the $ALE$ space
(see section 5 for more discussions). This allows to fix one of the constants
$k^{(1)}_\alpha,k^{(2)}_\alpha$. 
Therefore, only for $K3$ compactification with $SO(32)$ 
gauge group, the constants $k^{(1)}_\alpha,k^{(2)}_\alpha$ 
may be fixed (cf. the discussions in \afiuv). 
Besides, in the case of Abelian bundles one may impose additional 
equations, relating these numbers to the number of twisted matter fields
charged under these $U(1)$'s, which eventually fix these numbers
\gsw. However, in the case of non--Abelian instanton backgrounds, the
method which uses the supersymmetric index, seems to be more
restrictive. It allows us to fix the numbers 
$k^{(1)}_\alpha,k^{(2)}_\alpha$ {\it completely} (see also the tables 2,3
and 8).

\newsec{Gauge threshold corrections in N=1 $(0,2)$ orbifold 
compactifications}

A generic feature of gauge couplings in N=1,$d=4$ string vacua is their
dependence on scalars of chiral multiplets. The latter describe    
the universal dilaton $S$ and the moduli $T,U,V_i$ arising from the internal 
compactification.
At tree--level the gauge couplings of all gauge group $G_a$ factors
are given by the string--coupling, which 
is determined by the vev of the dilaton field $S$.
In effective string theory, this relation is modified\foot{A consistent
treatment is achieved by replacing the chiral multiplet $S$ with a linear 
multiplet \dfkz\nsz.}  by a mixing
between the dilaton $S$
and the other moduli, described by the non--harmonic function
$G^{(1)}$:
\eqn\mstr{
g_{\rm string}^{-2}={-iS+i\ov S \o 2}+{1\o 16\pi^2} G^{(1)}(T,U,V_i)\ .}
$G^{(1)}$ is the one--loop correction to the K{\"a}hler
potential \dfkz. In addition, one has to take
into account string threshold corrections originating from string modes
with masses above the string scale $M_{\rm string}$, which have been 
integrated out.
They (effectively) split the couplings at the string scale, i.e.
\eqn\gstring{
\lf.g_a^{-2}(\mu)\ri|_{\mu=M_{\rm string}}=k_ag_{\rm string}^{-2}+{1\o 16\pi^2}
\Delta_a\ .}
Whereas the K{\"a}hler potential $G$ receives contributions beyond
one--loop, which are (so far) not under control, the harmonic 
(Wilsonian) part of $\Delta_a$ is 
completely determined already at one--loop thanks to
non--renormalization theorems 
for the gauge kinetic function $f$. On the other hand, the
non--harmonic part of $\Delta_a$ is expected to have the opposite
sign of $G^{(1)}$. Thus with \mstr\ it has no influence on the
physical coupling \gstring\ -- at least at one--loop \dfkz\nsz.
However, it does affect the precise relation of $M_{\rm Planck}$ and
$M_{\rm string}$ at one--loop \nsz:
\eqn\precise{
M_{\rm Planck}^2=\lf[\im(S)+{1\o 16\pi^2} G^{(1)}(T,U,V_i)\ri]
M_{\rm string}^2\ .}

In this section we want to obtain generic results for the N=1 gauge
thresholds $\Delta_a$, focusing for concreteness, on N=1 toroidal
orbifolds \orbifolds. In these cases $\Delta_a$ receives only moduli--dependent 
contributions from N=2 subsectors \dklz.
Therefore, to obtain their analytic form, many results from the previous
sections may be borrowed. For concreteness, let us discuss six N=1 examples.

{\vbox{\ninepoint{
\def\ss#1{{\scriptstyle{#1}}}
$$
\vbox{\offinterlineskip\tabskip=0pt
\halign{\strut\vrule#
&~$#$~\hfil
&\vrule#
&~$#$~\hfil
&~$#$~\hfil
&~$#$~\hfil
&\vrule#
&~$#$~\hfil
&\vrule#
\cr
\noalign{\hrule}
&
\ 
&&
\ss{\rm orbifold}
&
\ss{\Gamma}
&
\ss{\tilde\Gamma}
&&
\ss{\rm \ perturbative\ gauge\ group}
&
\cr
\noalign{\hrule}
&
\ss{ I}
&&
\ss{ \IZ_2\times \IZ_2}
&
\ss{ (1,-1,0,0,0,0,0,0)}
&
\ss{ (0,0,0,0,0,0,0,0)}
&&
\ss{ E_6\times U(1)^2\times E_8'}
&
\cr
&
\ss{ II}
&&
\ss{ \IZ_2\times \IZ_2}
&
\ss{ (1,-1,0,0,0,0,0,0)}
&
\ss{ \tilde\Gamma_1=(0,0,0,0,0,0,0,0)\atop \tilde\Gamma_2=(2,0,0,0,0,0,0,0)}
&&
\ss{ E_6\times U(1)^2\times SO(16)'}
&
\cr
&
\ss{ III}
&&
\ss{ \IZ_2\times \IZ_2}
&
\ss{ (1,-1,0,0,0,0,0,0)}
&
\ss{ (2,0,0,0,0,0,0,0)}
&&
\ss{ E_6\times U(1)^2\times SO(8)'\times SO(8)'}
&
\cr
&
\ss{ IV}
&&
\ss{ \IZ_4}
&
\ss{ (1,-1,0,0,0,0,0,0)}
&
\ss{ (2,0,0,0,0,0,0,0)}
&&
\ss{ E_7\times U(1)\times SO(14)'\times U(1)'}
&
\cr
&
\ss{ V}
&&
\ss{ \IZ_6}
&
\ss{ (1,2,3,0,0,0,0,0)}
&
\ss{ (0,0,0,0,0,0,0,0)}
&&
\ss{ E_6\times U(1)^2\times E_8'}
&
\cr
&
\ss{ VI}
&&
\ss{ \IZ_6}
&
\ss{ (5,1,1,1,1,1,0,0)}
&
\ss{ (5,1,1,1,1,1,1,-1)}
&&
\ss{ SU(6)\times SU(3)\times SU(2)\times SU(8)'\times U(1)'}
&
\cr
\noalign{\hrule}}
\hrule}$$
\vskip-10pt
\noindent{\bf Table 5:}
{\sl Examples of N=1, $d=4$ heterotic $\IZ_\nu$ orbifolds with $(2,2)$ or 
$(0,2)$ world--sheet supersymmetry.}
\vskip10pt}}
The $\IZ_2\times\IZ_2$ models have the internal twists
$\theta_1=(-1,-1,+1), \ \theta_2=(-1,+1,-1)$, the $\IZ_4$ has 
$\theta={1 \o 4}(1,1,-2)$ and the $\IZ_6-II$
orbifold has the twist $\theta={1\o 6}(1,2,-3)$.
Models $(I),(V)$ have standard embeddings of the twist into the gauge
group, thus allowing for (2,2) world--sheet supersymmetry. 
On the other hand, models $(II),(III),(IV),(VI)$ are orbifolds with
non--standard twist
embeddings with only $(0,2)$ world--sheet supersymmetry.
Since only their N=2 sectors give rise to a modulus dependence of 
$\Delta_a$, let us investigate these sectors and give their relations
to the previous sections.

{\vbox{\ninepoint{
$$
\vbox{\offinterlineskip\tabskip=0pt
\halign{\strut\vrule#
&~$#$~\hfil
&\vrule#
&~$#$~\hfil
&\vrule#
&~$#$~\hfil
&\vrule#
&~$#$~\hfil
&\vrule#
\cr
\noalign{\hrule}
&
\ 
&&
{\rm 1st\  plane}\ \ \   T^1,U^1
&&
{\rm 2nd\  plane}\ \ \   T^2,U^2
&&
{\rm 3rd\  plane}\ \ \   T^3,U^3
&
\cr
\noalign{\hrule}
&
I
&&
\IZ_2st,\ n_1=-12,\nu_1=2
&&
\IZ_2st,\ n_2=-12,\nu_2=2
&&
\IZ_2st,\ n_3=-12,\nu_3=2
&
\cr
&
II
&&
\IZ_2nst (i),\ n_1=4,\nu_1=2
&&
\IZ_2nst (i),\ n_2=4,\nu_2=2
&&
\IZ_2st,\ n_3=-12,\nu_3=2
&
\cr
&
III
&&
\IZ_2nst (i), n_1=4,\nu_1=2
&&
\IZ_2nst (i), n_2=4,\nu_2=2
&&
\IZ_2nst (i), n_3=4,\nu_3=2
&
\cr
&
IV
&&
-
&&
-
&&
\IZ_2nst (i),\ n_3=4,\nu_3=2
&
\cr
&
V
&&
-
&&
\IZ_2st,\ n_2=-12,\nu_3=2
&&
\IZ_3st,\ n_3=-12,\nu_3=3
&
\cr
&
VI
&&
-
&&
\IZ_2nst (i),\ n_2=4,\nu_3=2
&&
\IZ_3nst (ii),\ n_3=6,\nu_3=3
&
\cr
\noalign{\hrule}}
\hrule}$$
\vskip-10pt
\centerline{\noindent{\bf Table 6:}
{\sl Twist invariant planes and $K3\times T^2$
details $(\nu_i,n_i)$ of the previous examples.}}
\vskip10pt}}
Here $\nu_i$ is the twist order of the N=2 subsector, which leaves
invariant the $i$th plane with moduli fields $T^i,U^i$. 
If one plane $i$ does not give rise to an N=2 sector, we just take
$\nu_i=0$ in all the following sums. In particular, this is the case
for $\nu=prime$.
Besides, in the cases $\nu_i/\nu\neq 1,\h$ the $U^i$--modulus is frozen.
The spectra of these models have been worked out in the appendix of
\janvadim. Also cancellation of anomalies, produced by triangle graphs
involving the K{\"a}hler and sigma--model connection, have been discussed
there. In particular, $G^{(1)}_{i,N=1}=0$, whenever $\nu_i/\nu=1,\h$,
however $G^{(1)}_{i,N=1}\neq 0$, if $\nu_i/\nu\neq 1,\h$.
Since the string--modes running in the loop arrange in N=2 multiplets
\dklz, the N=2 $\beta$--function coefficients of the underlying 
N=2 (sub)theory \betasieben\ will reappear 
in the calculations. The latter can be expressed by the N=1 K{\"a}hler
and sigma--model anomaly coefficients $\alpha^i_{G_A,G_{A'}}$ 
referring to the N=1 gauge group $G_A,G_{A'}$ \janvadim.

In total we get for the N=1 threshold corrections $\Delta_a$ (cf. \three) 
\eqn\THRESHONE{\eqalign{
\Delta_{G_A}&=-\sum_{i=1,2,3} {\nu_i\o \nu}\lf\{(12-6n_i)
\ln (\kappa T_2U_2)|\eta(T^i)\eta(U^i)|^4+k_A \sigma_{n_i}(T^i,U^i)+
k_AG^{(1)}_{i,N=2}\ri\}\cr
&\ \ \ \ \ -k_AG^{(1)}_{i,N=1}+const.\ ,\cr
\Delta_{G_{A'}}&=-\sum_{i=1,2,3} {\nu_i\o \nu}\lf\{(12+6n_i)
\ln (\kappa T_2U_2)|\eta(T^i)\eta(U^i)|^4+k_{A'} \sigma_{n_i}(T^i,U^i)+k_{A'}
G^{(1)}_{i,N=2}\ri\}\cr
&\ \ \ \ \ -k_{A'}G^{(1)}_{i,N=1}+const.\ .\cr}}
for vanishing Wilson lines $V_i$ ($s=0$).
With one Wilson--line $V:=V_{16}$ switched on $(s=1)$ we find:
\eqn\THRESHTWO{
\kern-1.5em\eqalign{
\Delta_{G_A}&=-\sum_{i=1,2,3} {\nu_i\o \nu}\lf\{{12-6n_i \o 12}
\ln (\kappa Y_i)^{12}|\chi_{12}|^2+k_A \sigma_{n_i}(T^i,U^i,V)+
k_AG^{(1)}_{i,N=2}\ri\}\cr
&\ \ \ \ \ -k_AG^{(1)}_{i,N=1}+const.\ ,\cr
\Delta_{G_{A'}}&=-\sum_{i=1,2,3} {\nu_i\o \nu}\lf\{{12+4n_i \o 12}
\ln (\kappa Y_i)^{12}|\chi_{12}|^2+
2n_i\ln\lf|{\triangle_{10}^{1/2}\o \chi_{12}^{5/12}}\ri|^2+
k_A \sigma_{n_i}(T^i,U^i,V)+k_AG^{(1)}_{i,N=2}\ri\}\cr
&\ \ \ \ \ -k_AG^{(1)}_{i,N=1}+const.\ .   \cr}}
The second piece in $\Delta_{G_{A'}}$ accounts for the subthreshold
effect which is caused by particles becoming massless for $V\rightarrow 0$.
In this case both $G_{A'}$ and the N=2 gauge group $G'$ are enhanced
(cf. also \betasieben\ and \Betas).

We see, that $\Delta_{G_A},\Delta_{G_{A'}}$ 
are given by $SO(2+s,2,\IZ)$ modular functions depending on the 
K{\"a}hler $T^i$ {\it and} complex structure moduli $U^i$ 
of the N=1 compactification and some topological data. The latter
are the instanton numbers $n_i$, which refer to the 
individual $N=2$ subsectors, described by $K3\times T^2$ dynamics
(cf. table 6).
It has already been stressed in \nsz, that, in contrast to
certain statements made in the past, the harmonic piece \sigm\ 
$\sigma_{n}$ in \THRESHONE\ and \THRESHTWO\ is of fundamental importance
to recover the correct decompactification limits to $d=6$ 
(cf. also \sixanomalypolynom) and $d=10$ dimensions (cf. next section). 

Our results \THRESHONE\ and \THRESHTWO\ hold quite general 
for N=1 orbifolds. In practice one only has to read the information
$n_i$ about their N=2 subsectors $\nu_i$ from the tables 2,3 and 8.
We may also go opposite: For given $n_i$, i.e. gauge twist 
$(\gamma,\tilde\gamma)$, construct an N=1 (modular invariant) $\IZ_\nu$
orbifold with twist $(\theta,{1\o\nu}\Gamma,{1\o\nu}\tilde\Gamma)$, 
whose N=2 subsector has the gauge twist ${1\o\nu_i}(\gamma,\tilde\gamma)$.
Modular invariance (see. e.g. \kawa) 
\eqn\modu{\eqalign{
&\sum_{I=1}^8\Gamma_I-\sum_{I=1}^8\tilde\Gamma_I=0\ \mod\ 2\cr
&\sum_{i=1}^3\theta_i^2-\sum_{I=1}^8
\Gamma_I^2-\sum_{I=1}^8\tilde\Gamma_I^2=0\ \mod\ 2\nu\cr}}
is quite restrictive and may rule out a lot of combinations.
Nevertheless, it is e.g. possible, to find N=1 orbifolds with $n_i=0$:
This happens, when the N=2 subsector, described by $K3$ dynamics has
an equal number of instantons in both $E_8$--factors, i.e. 
$(n^{(1)}_i,n^{(2)}_i)=(12,12)$.
In these cases, there is no one--loop correction to the 
gauge kinetic function\foot{This is also true for prime orbifolds, 
which do not possess any twist invariant planes.} (cf. section 5.1):
\eqn\exact{
f_{G_A,G_A'}=-iS+{\cal O}(e^{8\pi^2 i S})\ .}
The gauge group dependent part of the one--loop threshold correction
$b^{N=2}_{G}\triangle$ cancels against the group independent part $\sigma$.
These models look like N=4 models \solving, which do not have any 
perturbative corrections to  the
(two--derivative) gauge couplings due to the lack of 
enough fermionic contractions in a correlation function with  two gauge bosons.
However, our models have N=1 space--time supersymmetry
(with some N=2 subsector structure) and any two gauge boson correlator
must not vanish due to supersymmetry arguments.
On the other hand, these models do have moduli dependent 
wave--function renormalizations or one--loop corrections to the K\"ahler 
potential $G^{(1)}$. 
Let us give a list of these orbifolds, since they might 
be of some phenomenological use:
\vskip0.5cm
{\vbox{\ninepoint{
\def\ss#1{{\scriptstyle{#1}}}
$$
\vbox{\offinterlineskip\tabskip=0pt
\halign{\strut\vrule#
&~$#$~\hfil
&\vrule#
&~$#$~\hfil
&~$#$~\hfil
&\vrule#
&~$#$~\hfil
&\vrule#
\cr
\noalign{\hrule}
&
\ss{\IZ_\nu:\ \theta} 
&&
\ss{\Gamma}
&
\ss{\tilde\Gamma}
&&
\ss{\rm gauge\ group}
&
\cr
\noalign{\hrule}
&
\ss{\IZ_8-II:\ {1\o 8}(1,3,-4)}
&&
\ss{(1,1,0,0,0,0,0,0)}
&
\ss{(2,2,0,0,0,0,0,0)}
&&
\ss{U(1)\times E_7\times U(1)'\times E_7'}
&
\cr
&
\     
&&
\ss{(2,1,1,0,0,0,0,0)}
&
\ss{(2,0,0,0,0,0,0,0)}
&&
\ss{U(1)\times SU(2)\times E_6\times SO(16)'}
&
\cr
&
\   
&&
\ss{(3,1,1,1,1,1,0,0)}
&
\ss{(2,2,2,0,0,0,0,0)}
&&
\ss{U(1)^2\times SU(2)\times SO(10)\times SU(4)'\times SO(10)'}
&
\cr
&
\ss{\IZ_{12}-I:\ {1\o 12} (1,4,-5)}
&&
\ss{(3,1,1,1,1,1,0,0)}
&
\ss{(2,0,0,0,0,0,0,0)}
&&
\ss{U(1)^2\times SU(2)\times SO(10)\times SO(16)'}
&
\cr
&
\ss{\IZ_{12}-II:\ {1\o 12} (1,5,-6)}
&&
\ss{(3,2,1,0,0,0,0,0)}
&
\ss{(4,2,2,0,0,0,0,0)}
&&
\ss{U(1)^2\times E_6\times U(1)'\times SU(2)'\times E_6'}
&
\cr
&
\    
&&
\ss{(3,1,0,0,0,0,0,0)}
&
\ss{(2,0,0,0,0,0,0,0)}
&&
\ss{U(1)^2\times SO(12)\times SO(16)'}
&
\cr
&
\    
&&
\ss{\h(7,1,1,1,1,1,1,-1)}
&
\ss{(3,3,1,1,1,1,1,-1)}
&&
\ss{U(1)^2\times E_6\times SU(4)'\times SO(10)'}     
&
\cr
\noalign{\hrule}}
\hrule}$$
\vskip-10pt
\noindent{\bf Table 7:}
{\sl N=1 $\IZ_\nu$ orbifolds with vanishing perturbative corrections
to the gauge kinetic function~$f$:
Their twists $(\theta,{1\o \nu}\Gamma,{1\o \nu}\tilde\Gamma)$ 
and N=1 gauge groups.} 
\vskip10pt}}
The $\IZ_8-II$ models have an invariant plane for $\theta^2$ of
$\IZ_4$ (non--standard) $K3$--orbifold structure with instanton number $n_3=0$.
The one $\IZ_{12}-I$ example has a fixed plane for $\theta^3$, thus producing 
also a $\IZ_4$ $K3$--orbifold with $n_2=0$.
Finally, the three $\IZ_{12}-II$ cases have an invariant plane for $\theta^2$
of $\IZ_6$ $K3$--orbifold structure with $n_3=0$. Thus, 
there are no harmonic one--loop corrections to \gstring.

\newsec{M--theory origin of $d=4$ gauge couplings}

In this section we want to discuss the relation\foot{
Our $K3$ gauge threshold results (cf. sections 2 and 3) can also be 
related to $M$--theory compactified on $S^1/\IZ_2\times K3$
(see also \wstrong\wyll).} of our N=1 gauge threshold results
\THRESHONE\ and \THRESHTWO\ 
to the strongly coupled heterotic string in ten dimensions, which
is described by M--theory compactified on $S^1/\IZ_2$ \hw.
This question has been raised in \bd\ and worked out for
standard--embedding in \nsz.

\subsec{Gauge kinetic function in N=1, $d=4$ weakly coupled
heterotic string theory} 

The relevant object to link the four--dimensional one--loop
corrections to the strong coupling expansion of $M$--theory is 
the gauge kinetic function $f$ of the gauge groups $G_A,G_{A'}$, 
in which the findings of the previous sections are summarized:
\eqn\ffunction{\kern-2em
f_{G_A,G_{A'}}(S,T^i,U^i,V)=-iS+\kern-0.7em\sum_{i=1,2,3}{\nu_i\o \nu}\lf\{
{1\o 5}(\p_{T^i}\p_{U^i}-{1\o 4}\p_V^2)h_{n_i}-{1\o 5\pi^2}
\ln\Psi_{n_i}(T^i,U^i,V)\ri\}+{\cal O}(e^{2\pi i S})\ .}
After the rescalings $S\rightarrow 4\pi S,\ f\rightarrow 4\pi f$, which 
corresponds to the dilaton choice 
$S={\theta_a\o 2\pi}+i{4\pi \o g^2_{\rm string}}$
the large $T^i$--expansion\foot{For $n_i=-12$ and restricting the sum
to $i=3$, 
we recover the results of \nsz.} (large K\"ahler moduli)
of \ffunction\ becomes (using $\Delta_{35}
\rightarrow e^{4\pi iT},\ \Delta_{10}\rightarrow e^{2\pi iT}$):
\eqn\red{\eqalign{
f_{G_A}&=-iS-i\sum_{i=1,2,3}{\nu_i\o \nu}\lf({b^{N=2}_{G_{i}}\o 12}-1\ri)T^i
+{\cal O}(e^{8\pi^2 i S})=
-iS+i\sum_{i=1,2,3}{\nu_i\o \nu}\ {n_i\o 2}T^i+{\cal O}(e^{8\pi^2 i S})\ ,\cr 
f_{G_{A'}}&=-iS-i\sum_{i=1,2,3}{\nu_i\o \nu}
\lf({b^{N=2}_{G'_{i}}\o 12}-1\ri)T^i+{\cal O}(e^{8\pi^2 i S})=
-iS-i\sum_{i=1,2,3}{\nu_i\o \nu}\ {n_i\o 2}T^i+{\cal O}(e^{8\pi^2 i S})\ .\cr}}
These expressions may be directly identified 
with the $f$--functions, which arise upon dimensional reduction of the 
weekly--coupled ten--dimensional heterotic string.
This holds --at least in this limit-- for generic $n$, as
in this reduction $n$ enters only as the instanton number 
of the gauge bundle in the Bianchi identity.
Therefore, from the ten--dimensional viewpoint, the form of the 
gauge kinetic function \ffunction\ in four dimensions is dictated by
the Green--Schwarz anomaly cancellation in ten dimensions, together
with target--space duality \ib\nsz.
As a remark, let us mention that for the compactifications
we have considered, i.e. N=1 orbifolds with N=2 sectors, which are
described by $K3\times T^2$ dynamics, \ffunction\ can be 
also deduced from the relevant Green--Schwarz anomaly cancellation
terms in six dimensions \sixanomalypolynom, since in \red\ each N=2
subsector may be thought as a decompactification limit $T^i\ra\infty$
to six dimensions. 

For (2,2) Calabi--Yau compactifications $X$, there exists a relation
of one--loop gauge threshold corrections 
to the (large K\"ahler modulus expansion thereof) topological index $F_1$ \cecotti.
The identity $\Delta_{E_6}-\Delta_{E_8'}\lra 12F_1$ allows us, to write
for the large radius expansion of 
the gauge kinetic functions \vk:
\eqn\holo{
f_{E_8'}-f_{E_6}\longrightarrow 2\sum_{i=1}^{h_{(1,1)}} 
t^i\int_X J_i\wedge c_2(R)\ .}
Here $J_i$ is a basis for the K{\"a}hler class $H^{(1,1)}$ and $c_2$ is
the second Chern class of the Calabi--Yau threefold $X$.
The same limit \holo\ appears after a dimensional reduction of the
ten--dimensional Green--Schwarz term specializing to the difference
of the $E_8,\ E_6$ axionic couplings.
Applied\foot{The relation \holo\ keeps its validity in the orbifold limit,
although some CY moduli $t^i$ are frozen at finite values in the K{\"a}hler 
moduli space.} to the orbifolds with standard--embedding 
(e.g. $I$ and $V$ of table 5), we get e.g. 
\eqn\workout{
f_{E_8'}-f_{E_6}\longrightarrow -\h\chi_{K3}\sum_{i=1,2,3}
{\nu_i\o \nu}T^i=-{1\o 4}\chi(T^6/\IZ_\nu)\sum_{i=1,2,3}' t^i \ ,}
with the CY moduli $t^i={\nu_i\o \nu}T^i$ and the CY--Euler 
number\foot{For all toroidal $\IZ_\nu$ orbifolds $X$ we have
$\chi(T^6/\IZ_\nu)=48$, except $\chi(T^6/\IZ_3)=72$. However a $\IZ_3$
orbifold does
not give rise to moduli dependent thresholds $\nu_i=0$ and we may
safely introduce $\chi(X)=48$ in \workout. }
$\chi(X)=48$ of the underlying N=1 orbifold $X$. The prime means, that
we only sum over such moduli, which appear from N=2 subsectors.
This limit is in agreement with \red\ for $n_i=-12$. 
In that case it is straightforward to work out the integral (cf. below for 
the more general case).

\subsec{M--theory on $S^1/\IZ_2\times T^6/\IZ_\nu$}

In \bd\nsz\ it was argued that \ffunction\ encodes for standard--embedding
the strong coupling expansion (an expansion in the eleven dimensional 
gravitational coupling constant $\kappa^2:=\kappa_{11}^2$)
of M--theory on $S^1/\IZ_2$ compactified on a CYM $X$.
I.e. a perturbative heterotic gauge threshold calculation (as performed in the
previous section and summarized in eq. \ffunction) gives the 
gauge couplings of M--theory on $S^1/\IZ_2$ compactified on this CYM. 
To zeroth order in $\kappa^2$ the relative sizes of the CY and $S^1$ are 
not relevant and the expansion of the 
strongly coupled heterotic string theory gives the same effective action in 
four dimensions as the dimensional reduction of the weekly coupled 
ten--dimensional heterotic string. 
Moreover, at higher orders in $\kappa$ their four--dimensional 
effective actions take the same analytic form and thus cannot be 
distinguished from each other. 
In this section we want to discuss the case with non--standard embedding, 
since it leads to realistic string vacua \wend. We will see, how 
\red\ arises from $M$--theory compactification.

The $G_A,G_{A'}$ gauge fields live on the two nine--branes.
After compactification on the CYM $X$ their coupling is given by
\eqn\coupling{
g^{-2}_{G_A,G_{A'}}={  2V_X \o (4\pi )^{5/3}\kappa^{4/3}}\ ,}
to order $\kappa^{2/3}$ relative to the bulk. Here, $V_X$ is the 
Calabi--Yau volume at the boundaries $x_{11}=0$ and $x_{11}=\pi\rho$, 
respectively.
Corrections coming from interactions to the bulk, start at order $\kappa^{4/3}$
and modify this relation. 
This results in a variation of the CY volume $V$ over the interval $x_{11}$.
Therefore, to determine the two gauge couplings $g_{G_A}$ and $g_{G_{A'}}$ 
we need an expression for the two volumina $V(0)$ and $V(\pi\rho)$ 
at the two fixed points.
Here $\rho$ is the radius of the eleventh dimension $S^1$. In Wittens linear 
approximation their difference is given by \wstrong
\eqn\linear{
V(\pi\rho)-V(0)=2\pi^2 \rho \lf({\kappa\o 4\pi}\ri)^{2/3}\int_X \omega\wedge
{\tr F\wedge F-\h \tr R\wedge R \o 8\pi^2}\ .}
The r.h.s. is an integral over the CYM $X$ to be worked out at the boundary
$x_{11}=0$. In particular this means, that the gauge fields 
(in the following denoted by $F^{(1)}$) refer to the gauge group $G_A$, 
which lives on the wall $x_{11}=0$.
Fortunately, the r.h.s. is independent on $x_{11}$ in the linear approximation.
This means that the CY moduli entering there are 
$x_{11}$--independent and the whole integral describes a generic 
topological coupling on the CYM $X$. Nonetheless, the interpretation
of the CY moduli appearing in the four--dimensional low--energy effective 
action as chiral fields is different for $M$--theory on $S^1/\IZ_2\times X$ 
and $10d$ heterotic string on $X$. Since the former are coming from an 
eleven-dimensional
theory, they describe five--dimensional fields, which have to be averaged.
To the order we are considering, this averaging means, that all CY
moduli fields (more precisely: their non--axionic parts)
refer to metric scalars $g_{MN}^{11d}$ in the middle $x_{11}=\h\pi\rho$ of 
$S^1/\IZ_2$. More details about this identification 
can be found in \hpn\andre. 
Besides, further aspects have been analyzed in a burst of recent
papers \hpn\andre\all, 
(however, all of them dealing with standard embedding).

We want to work out\foot{Recently, the relation
eq. \linear\ has been discussed in \benakli\ for one specific CY
example and in \choi\ with the emphazise on the axionic symmetries of the
K\"ahler moduli $t^i$.} \linear\ for the models we have considered in
section 4, i.e. in particular for instanton non--standard embeddings.
Since we are compactifying on a Calabi--Yau manifold,
$R_{ik}=0$ and $F^{(a)}_{ij}=0$, and 
the only non--vanishing components of $\tr R\wedge R$ and $\tr F\wedge F$ 
come from the combinations $R_{i\ov jk\ov l}$ and $F_{i\ov j}F_{k\ov l}$, 
respectively \GSW, we may expand the fields 
\eqn\expand{\eqalign{
\tr R\wedge R&=\chi_i d^i\ \ , \ \ \chi_i=\int_{C_i}\tr R\wedge R\ ,\cr
\tr F^{(1)}\wedge F^{(1)}&=n^{(1)}_i d^i\ \ ,\ \ n^{(1)}_i=
\int_{C_i}\tr F^{(1)}\wedge F^{(1)}\cr}}
w.r.t. a basis $d^i$ of harmonic $(2,2)$--forms ($i=1,\ldots,h_{(2,2)}$).
Of course, from Poincar{\'e} duality $h_{(2,2)}=h_{(1,1)}$.
The corresponding $4$--cycles $C_i$ are chosen to fulfil the following 
intersection properties
\eqn\cycle{
\int_{C_i}d^j=\delta_i^j\ \ ,\ \ \int_X J_i\wedge d^j=\delta_i^j\ ,}
where $J_i$ is a basis of $(1,1)$--forms to which the K{\"a}hler moduli $t^i$
($i=1,\ldots,h_{(1,1)}$) are associated.
Besides we expand the K{\"a}hler form $\omega$ of $X$
$\omega=t^iJ_i$. 
In the orbifold limit of a CYM all the instantons are stuck at the
fixed points. 
Therefore, the $4$--cycle integrals \expand\ receive contributions
only from orbifold singularities (of (complex) codimension $2$), 
since these are the only sources for curvature and places for 
non--trivial gauge connections.  
Thus we have to consider all $4$--cycle $C_{f_\alpha}$ integrals
around fixed points $f_\alpha$ of codimension $2$ 
rather\foot{Since prime orbifolds ($\nu=prime$) have no 
codimension $2$ fixed points, the integral in \linear\
vanishes. In those cases, there are also no perturbative one--loop 
corrections, i.e. $\nu_i=0$ in (4.4) and (4.5).} than $3$.
These are precisely the N=2 sectors of the N=1 
orbifold under consideration and they can be described by $K3$ dynamics.
Locally at these points, the manifold is replaced by an $ALE$--space 
with $A_\nu$--type singularity.
These are asymptotic locally flat non--compact spaces. 
They have the Euler number $\chi_{ALE_\nu}={\nu^2-1 \o \nu}$ \cand.
Thus, in \expand\ we obtain the coefficients:
\eqn\zerosources{\eqalign{
n^{(1)}_\alpha&=\int_{{\Cc}_{f_\alpha}}\tr F^{(1)}\wedge F^{(1)}=
k_\alpha^{(1)}+\h{1\o \nu_\alpha}\sum_{I=1}^8\gamma_\alpha^I
(\nu_\alpha-\gamma^I_\alpha)\ , \cr
\chi_{ALE_{\nu_\alpha}}&=\int_{{\Cc}_{f_\alpha}}\tr R\wedge R=
{\nu_\alpha^2-1 \o\nu_\alpha}\ .\cr}}
Here $n^{(1)}_\alpha$ is the individual gauge instanton number \index\
at the fixed point
$f_\alpha$ (which is supposed to have twist order $\nu_\alpha$) 
and the curvature singularity contributes $\chi_{ALE_{\nu_\alpha}}$ to the 
gravitational instanton contribution.
Let us mention the identity $\int_{C_{f_\alpha}}
dH=n^{(1)}_\alpha+n^{(2)}_\alpha+\chi_{ALE_{\nu_\alpha}}=0$, 
which expresses local charge cancellation at the fixed point $f_\alpha$.
In addition, after \cycle, to get a non--zero wedge product, 
the K{\"a}hler $J_i$ form has to lie in the remaining orthogonal 
one (complex) dimensional plane.
This is the plane $T_i^2$, left invariant under the orbifold twist, with
K{\"a}hler modulus $t^i\equiv {\nu_i\o\nu}T^i$. 

In total, summing up all source contributions \zerosources\ 
at codimension $2$ fixed points and noting the fact (which holds for every
$t^i$, which has a set of codimension 2 fixed points $f_\alpha$)
\eqn\eulerk{
\sum_k N_k \chi_{ALE_k}=\chi_{K3}=n^{(1)}_i+n^{(2)}_i\ ,}
where $N_k$ is the number\foot{
Concretely: $\nu=2$: $T^4/\IZ_2$,\  $N_2=16$;\  
            $\nu=3$: $T^4/\IZ_3$,\  $N_3=9$;\ 
            $\nu=4$: $T^4/\IZ_4$,\  $N_4=4,\ N_2=6$;\  
            $\nu=6$: $T^4/\IZ_6$,\  $N_6=1,\ N_3=4,\ N_2=5$.} of 
(dimension 1) fix--planes
of order $k$, we derive\foot{Of course, for the orbifold examples in
table 7 we get $V(\pi\rho)-V(0)=0$. 
This allows for equal couplings at both boundaries.} for \linear:
\eqn\linearo{\eqalign{
V(\pi\rho)-V(0)&=2\pi^2 \rho \lf({\kappa\o 4\pi}\ri)^{2/3} 
\sum_i{\nu_i\o\nu}T^i[n^{(1)}_i-\h(n^{(1)}_i+n^{(2)}_i)]\cr
&=-2\pi^2 \rho \lf({\kappa\o 4\pi}\ri)^{2/3}\sum_i{\nu_i\o\nu}n_i T^i\ .\cr}}

To compare with \red\ we have to translate the eleven dimensional
scales $\kappa_{11}$ and $R_{11}\equiv \pi\rho$ 
to ten dimensional heterotic string quantities \bd:
\eqn\trans{
{\rho\o \kappa_{11}^{2/3}}={1\o 2^{7/3}\pi^{8/3}}{1\o \alpha'}\ .}
With \coupling\ and \linearo, the difference of the two gauge couplings 
$\alpha_{G_A}$ and $\alpha_{G_{A'}}$ becomes 
\eqn\final{
g^{-2}_{G_{A'}}-g^{-2}_{G_A}=-{1\o 32\pi^3}{1\o\alpha'}\sum_i{\nu_i\o\nu}
n_i T^i\ ,} 
which agrees with \red\ up to numerical constants.
This leads to the generalization of \holo\ to non--standard embedding 
orbifolds:
\eqn\orbifoldholo{
f_{G_{A'}}-f_{G_A}\longrightarrow -\sum_i{\nu_i\o\nu}n_i T^i\ .} 
Tracing back \orbifoldholo\ to its origin \linear\ we conjecture 
the large radius expansion of the holomorphic index
\holo\ for $(0,2)$--compactifications:
\eqn\hologeneral{
f_{G_{A'}}-f_{G_A}\longrightarrow 2\sum_{i=1}^{h_{(1,1)}} 
t^i\int_X J_i\wedge \lf(\tr F\wedge F-\h \tr R\wedge R\ri)\ .}

Let us make some final remarks: 
The techniques, developed in this section for the dimensional reduction
on (0,2) orbifolds, may be also used to extract other 
(than harmonic gauge coupling) terms in the four dimensional effective action.
In particular, we find it interesting to trace back the origin of the 
non--harmonic coupling $G^{(1)}$, appearing in eqs. 
\THRESHONE\ and \THRESHTWO\ to ten or eleven dimensions.
On the other hand, in four dimensions it is due to 
IR--effects and its large radius behaviour (maximally $log T$)
is quite different than that of the gauge couplings. 
Therefore, we cannot obtain it in the limit described above, which gives 
order $T$ effects in the effective $4d$ action.
Moreover, it would be interesting to determine the 
expression \hologeneral\ (and eventually eqs. \THRESHONE\ and \THRESHTWO) 
from $F$--theory on a fourfold by considering $7$--brane exchange 
interactions.
A similar treatment has been accomplished in $d=8$ with $F$--theory on $K3$
\ls. 
There, four--point couplings $R^4$ and $F^4$ could be calculated
by means of $7$--brane exchanges. 
However, the $d=4$ case is more involved because of the complicated 
bundle structure on the fourfold and $7$--branes.

\subsec{Including NS $5$--branes}

In \dmw, the possibility of adding NS $5$--branes into
the space--time was considered.
This then may be considered as additional source term in the Bianchi
identity for  the $4$--form $G$ \wstrong.
Again, this effect may be studied in N=1 non--perturbative orbifold 
constructions \afiuv. These orbifolds are non--modular invariant
at the perturbative level. That means that they have gauge and/or 
gravitational anomalies at the perturbative level. However,
non--perturbative effects, like additional $5$--branes \sw\dmw, render the
theories consistent. 
Usually, they may have N=2 subsectors, whose non--perturbative
formulation may be traced back\foot{Provided there are enough massless blow 
up modes.} to known N=1, $d=6$ smooth $K3$ dynamics of 
tensionless strings, small instantons or $5$--branes, 
compactified on the torus $T^2$.
In M--theory, where these effects are described by NS 5--branes
approaching one of the 9--branes, the characteristic length is their relative 
distance $<\Phi_i>$ to one 9--brane. 
The field $\Phi$ is a real scalar of a tensor multiplet in 
$d=6$, N=1. After 
torus compactification it becomes  a scalar of an N=2, $d=4$ vector multiplet,
whose gauge field show the coupling \ganor
\eqn\treetensor{
\re(U)F_{\mu\nu}F^{\mu\nu}\ .}
Physical quantities are expanded w.r.t. 
$1/u\sim e^{2\pi^2 i (T_1+i<\Phi>T_2)}$,
accounting for the instantons, which are strings of tension $<\Phi>$,
wrapped around the torus $T^2$. 
In particular, this gives e.g. SW--like expansions (in $u$) 
for the gauge coupling \treetensor.
Together with the usual (conventional) perturbative
expansion\foot{I.e. 
$S\rightarrow\infty$, which is clearly not the right limit, when one wants to 
take into account effects of the NS 5--branes. 
This limit puts together the two 
9--branes and we lose the effects of the 5--branes, which were in between 
them.}  w.r.t. the dilaton $S$, in $d=4$ we are then left with two
expansions, valid in different regimes of the moduli space. 
On the other hand, the r{\^o}le of a perturbative string threshold correction
in the sense of sects. 2 and 3
as a perturbative expansion in the dilaton field $S$
does no longer make sense, since we are dealing with an 
anomalous or non--modular invariant theory.
The perturbative part of the partition function alone is not sufficient 
to consider, 
since it is not one--loop modular invariant and we do not know its 
non--perturbative extension rendering modular invariance.
However, modular invariance is a 
quite important ingredient in string--perturbation theory. In fact, the 
results of sections 2-4 heavily rely on the modular invariance of the 
partition function and we are not allowed to apply them for those
kinds of models,
although a naive guess might urge us, to just insert in \Kth\ 
instanton numbers $(n_i^{(1)},n_i^{(2)})$, fulfilling 
$n_i^{(1)}+n_i^{(2)}\neq 24$. 
In \red\ this would lead to an asymmetric $T$--dependence.

However, for the twisted sectors of the N=1 non--perturbative
orbifold we do not have any description.
Less is known about the non--perturbative effects, which are supposed
to reinforce modular invariance or anomaly freedom.
It is believed that the analog of small instanton dynamics in N=1,
$d=6$ is played by chirality change in N=1,$d=4$ \silver. 

Let us draw one conclusion (just from considering the N=2 subsectors
of the non--perturbative orbifolds):
The coupling \treetensor\ shows a quite different structure than what
one expects in ordinary string perturbation theory, where the dilaton
$S$ contr{\^o}ls all tree--level couplings \mstr. 
We have different expansions for the gauge couplings, valid in different 
regions of the moduli space. So far lacking a complete (non--perturbative) 
heterotic description, which eventually puts $S$ and the moduli $T,U,V$ on 
the same footing.
This is naturally provided by $F$--theory compactifications, which
will certainly lead 
to quite new concepts in string phenomenology \klF.

\ \ 
{\bf Acknowledgements:}
I wish to thank G.L. Cardoso, J.--P. Derendinger, and K. Intriligator
for interesting discussions. 
Moreover, I thank Z. Lalak, W. Lerche, H.P. Nilles, B. Pioline, 
and especially P. Mayr for helpful discussions.
The major part of this work was carried out at Neuch{\^a}tel University with
the support of the
Swiss National Science Foundation, the European Commission TMR programme 
ERBFMRX--CT96--0045, and OFES no. 95.0856.

\goodbreak

\appendix{\appA}{Jacobi functions}

The coupling of the genus \genusB\ to one Wilson line in \Z\ is decribed
Jacobi forms \ke\kz.
A Jacobi form (for more details see \zagier) $f_{s,m}$ of weight
$s$ and index $m$ enjoys
\eqn\enjoy{\eqalign{
f_{s,m}\lf({a\tau +b \o c\tau+d},{z\o c\tau+d}\ri)&=(c\tau+d)^s
e^{\pi i {m c z^2 \o c\tau+d}}f_{s,m}(\tau,z)\ ,\cr
f_{s,m}(\tau,z+\lambda\tau+\mu)&=e^{-\pi im(\lambda^2\tau+2\lambda z)}
f_{s,m}(\tau,z)\ ,\cr}}
for $\lf(\matrix{a&b\cr c& d\cr}\ri)\in SL(2,\IZ)$ and $\lambda,\mu\in\IZ$.
Prominent examples (for index 1) are the Jacobi $\theta$--functions 
($y=e^{2\pi i z}$):
\eqn\jac{
\theta\lf[\alpha\atop\beta\ri](q,y)=\sum_{n\in \IZ}
q^{\h(n+\h \alpha)^2}e^{\pi i(n+\h\alpha)\beta}\ y^{n+\h \alpha}\ .}
Explicitly,
\eqn\jacobi{\eqalign{
\theta\lf[1 \atop 1\ri](q,y)\equiv\theta_1(q,y)&=i\sum_{n \in
\IZ}(-1)^nq^{\h(n+\h)^2}y^{n+\h}\cr
\theta\lf[1\atop 0\ri](q,y)\equiv\theta_2(q,y)&=\sum_{n \in 
\IZ}q^{\h(n+\h)^2}y^{n+\h}\cr
\theta\lf[0\atop 0\ri](q,y)\equiv\theta_3(q,y)&=\sum_{n \in 
\IZ}q^{\h n^2}y^n\cr
\theta\lf[0\atop 1\ri](q,y)\equiv\theta_4(q,y)&=\sum_{n \in 
\IZ}(-1)^nq^{\h n^2}y^n\ .\cr}}
The ring of Jacobi forms of index $1$ is generated by the
Jacobi--Eisenstein
functions
\eqn\eisenhat{\eqalign{
E_{4,1}(q,y)&=\h[\theta_2(q,y)^2\theta_2^6+
\theta_3(q,y)^2\theta_3^6+\theta_4(q,y)^2\theta_4^6]\ ,\cr
E_{6,1}(q,y)&=\h[\theta_4^6\theta_4(q,y)^2(\theta_2^4+\theta_3^4)+
\theta_3^6\theta_3(q,y)^2(\theta_4^4-\theta_2^4)-
\theta_2^6\theta_2(q,y)^2(\theta_3^4+\theta_4^4)]\ ,\cr}}
with $\theta_1=\theta\lf[1\atop 1\ri],\ 
\theta_2=\theta\lf[1\atop 0\ri],\ 
\theta_3=\theta\lf[0\atop 0\ri]$ and $\theta_4=\theta\lf[0\atop 1\ri]$.
The $K3$ elliptic genus \japan\ is a Jacobi form of weight $0$ and
index $1$:
\eqn\kdrei{
{\cal Z}_{K3}(q,y)={1\o 72}{E_4^2E_{4,1}(q,y)-E_6E_{6,1}(q,y)\o \eta^{24}}\ .}
Any Jacobi form $f_{s,1}(q,y)$ of index $1$ can be decomposed as
\eqn\deco{
f_{s,1}(q,y)=f_{s,1}^{even}(q)\theta_{even}(q,y)
+f_{s,1}^{odd}(q)\theta_{odd}(q,y)\ ,}
with $\theta_{even}(\tau,z)=\theta_3(2\tau,2z)$ and 
$\theta_{odd}(\tau,z)=\theta_2(2\tau,2z)$.
Finally, we define
\eqn\DEF{
\hat{f}_{s,1}(q)=f_{s,1}^{even}(q)+f_{s,1}^{odd}(q)\ .}

Applying \deco\ for $E_{4,1}(q,y)$ provides (for $y=1,z=0$) the decomposition
of one $E_8$ gauge factor into $E_7\times A_1$
\eqn\DECO{
E_4=E_{4,1}(q,1)=E_{4,1}^{even}(q)\theta_3(2\tau)
+E_{4,1}^{odd}(\tau)\theta_2(2\tau)\ ,}
with  the $E_7$--characters
$Z_{\bf 133}=E_{4,1}^{even}$ and $Z_{\bf 56}=E_{4,1}^{odd}$ 
and $A_1$--characters $Z_{A_1^{\bf 0}}=\theta_3(2\tau)$ and 
$Z_{A_1^{\bf 1}}=\theta_2(2\tau)$.
Then, the coupling of one Wilson to the Narain
lattice sum $Z_{3,2}$ and to one $E_8$--gauge group factor (or more
generally to the remaing part of the index) 
can be described by the Jacobi form  $E_{4,1}$ (or $f_{s,1}$)
\eqn\COUP{\kern-1.5em\eqalign{
f_{s,1}(q,y)\otimes Z_{3,2}(q,\ov q)&:=  \sum_{k=even}
\sum_{m_i,n^i} 
e^{2\pi i\tau({1\o 4} k^2+m_1n^1+m_2n^2)}\ 
e^{-2\pi\tau_2 |p_R|^2}\ f_{s,1}^{even}(q)\cr
&+\sum_{k=odd}
\sum_{m_i,n^i} 
e^{2\pi i\tau({1\o 4} k^2+m_1n^1+m_2n^2)}\ 
e^{-2\pi\tau_2 |p_R|^2}\ f_{s,1}^{odd}(q)\ ,\cr}}
with $p_R$ defined in eq. \Z.                           

\appendix{\appB}{Orbifold details}

In this section we give the phase matrix $k_{(a,b)}$, which appears in
the $K3$ index \char.

\subsec{$\IZ_2$ orbifold}
\eqn\kii{
k_{(a,b)}=64\pmatrix{0&1\cr
                     1&e^{-\pi i {1\o \nu^2}(2-\gamma^2)}}}
\subsec{$\IZ_3$ orbifold}
\eqn\kiii{
k_{(a,b)}=36\pmatrix{0&1&1\cr
1& e^{-{1\o 9} \pi i (2-\gamma^2)}  & e^{-{2\o 9}\pi i  (2-\gamma^2)}  \cr
1& e^{-{5\o 9} \pi i (2-\gamma^2)}  & e^{-{4\o 9} \pi i (2-\gamma^2)}   }}
\subsec{$\IZ_4$ orbifold}
\eqn\kiiii{
k_{(a,b)}=16\pmatrix{0&1&4&1\cr
 1& e^{-\pi i {1\o 16}(2-\gamma^2)}&e^{-\pi i {1\o 8}(2-\gamma^2)}&
                               e^{-\pi i {3\o 16}(2-\gamma^2)}     \cr
                     4& e^{\pi i {3\o 8}(2-\gamma^2)} & 
4e^{-\pi i {1\o 4}(2-\gamma^2)}& e^{\pi i {1\o 8}(2-\gamma^2)}     \cr
                     1& e^{\pi i {9\o 16}(2-\gamma^2)}& 
e^{\pi i {1\o 8}(2-\gamma^2)}& e^{-\pi i {9\o 16}(2-\gamma^2)}  }}

\subsec{$\IZ_6$ orbifold}
\def\Ga{\Gamma}
\eqn\kii{
k_{(a,b)}=4\pmatrix{ 0&1&9&16&9&1 \cr
1&e^{{1\o 3}\pi i\Ga} & e^{{2\o 3}\pi i\Ga} &e^{\pi i\Ga}&e^{{4\o
3}\pi i\Ga}&e^{{5\o 3}\pi i\Ga}\cr
9&e^{{2\o 3}\pi i\Ga} & 9e^{{4\o 3}\pi i\Ga} &1&9e^{-{4\o
3}\pi i\Ga}&e^{-{2\o 3}\pi i\Ga}\cr               
16&e^{\pi i\Ga} & 1 &16e^{\pi i\Ga}&1&e^{-\pi i\Ga}\cr
9&e^{-{2\o 3}\pi i\Ga} & 9e^{{2\o 3}\pi i\Ga} &1&9e^{{4\o
3}\pi i\Ga}&e^{-{4\o 3}\pi i\Ga}\cr
1&e^{-{1\o 3}\pi i\Ga} & e^{-{2\o 3}\pi i\Ga} &e^{-\pi i\Ga}&e^{-{4\o
3}\pi i\Ga}&e^{{1\o 3}\pi i\Ga}\cr}                   }
with $12\Gamma:=\gamma^2+\tilde\gamma^2-2$.

\appendix{\appC}{Lowest Expansion of $B_A$}

In this part we derive the lowest $V$--expansion in \productz.
We consider the integral \vadim\ with the integrand \KthB\
w.r.t. to a gauge group $G_A$, which is not enhanced
at any point in the moduli space, except $T\ra i\infty$.
It is easy to show
\eqn\zero{
\lf. {\p\Delta_A \o \p V}\ri|_{V=0}=0\ ,}
due to a possible relabelling of the quantum numbers appearing
in the sum of $Z$.
We use the identity $(p_{R,0}:=\lf.p_R\ri|_{V=0})$
\eqn\fz{\hskip-0.5cm
\lf.{\p^2\o \p V^2} Z_{3,2} \right|_{V=0}=
{{2i}\o {-iU+i\ov U}} {{\p}\o {\p T}} Z_{2,2}
+{{2i}\o {-iT+i\ov T}} {{\p}\o {\p U}} Z_{2,2}
+{{8\pi^2\tau_2^2}\o {(-i T+i \ov T)(-i U+i \ov U)}}\bar p_{R,0}^2
\lf(k^2-{1\o{2\pi \tau_2}}\ri)Z_{2,2}}
to perform the integrand of \vadim\ leading to
\eqn\fd{\eqalign{
\lf.{{\p^2}\o{\p V^2}}\Delta_A\ri|_{V=0}&=
{{2i}\o{-i U+i\ov U}} 
{{\p}\o {\p T}}\tilde\Delta_A+{{2i}\o{-i T+i\ov T}}
{{\p}\o{\p U}}\tilde\Delta_A\cr
&+\int {{d^2\tau}\o{\tau_2}} 
\sum  {{8\pi^2\tau_2^2}\o{(-i T+i 
\ov T)(-i U+i \ov U)}}Z_{2,2}
\bar p_{R,0}^2 \lf(k^2-{{1}\o{2\pi \tau_2}}\ri) \cr}}
Here $\tilde\Delta_A$ denotes the threshold correction for vanishing Wilson 
line. Let us denote the last integral by $R$.
Then we perform a manipulation similar to the one introduced in \agntz.
With
\eqn\fv
{(-iU+i\ov U)^2 \p_{\ov U}\lf[\sum{{\ov p^2_{R,0}}\o{(-iU+i\ov U)}} 
q^{\h|p_L|^2}\bar q^{\h|p_R|^2}\ri]={{\ov T-T}\o{\pi \tau_2}}\ 
\sum \p_{\bar \tau}[\tau_2 \p_T \Zc]\ ,}
we calculate $\p_{\bar U} R$
\eqn\ff{\eqalign{
\p_{\ov U}R
&={{-8\pi i}\o{(-i U+i\ov U)^2}}\int
{{d^2\tau}\o{\tau_2}}
\sum_k\tau_2\p_{\bar \tau}(\tau_2 \p_T {\Zc}_0) \lf(k^2-{{1}\o{2\pi
\tau_2}}\ri){\Cc}(\tau)\cr
&={{-2}\o{(-iU+i\ov U)^2}} \ {{\p\tilde\Delta_A}\o{\p T}}\cr}}
where the last eq. follows after a partial integration.
After a duality respecting integration we find:
\eqn\fs{
R=-2{{\p\tilde\Delta_A}\o{\p U}}{{\p\tilde\Delta_A}
\o{\p T}}\ .}
Therefore we have
\eqn\fsi{
\lf.{{\p^2 \Delta_A}\o{\p V^2}}\ri|_{V=0}={{2}\o{\ov U-U}} 
\tilde\Delta_{a,T}
+{{2}\o{\ov T-T}} \tilde\Delta_{A,U}-2\tilde \Delta_{A,U}
\tilde \Delta_{A,T}}
and finally:
\eqn\a{
\lf.{{\p^2 \Delta_A}\o{\p
V^2}}\ri|_{V=0}=
{{8\eta'(T)\eta'(U)}\o{\eta(T)\eta(U)}}-{{2}\o{(T-\ov T)(U-\ov U)}}\ .}
A similar result is obtained for 
$\lf.{{\p \Delta_A}\o{\p \bar V^2}}\ri|_{V=0}$.
However one may show
\eqn\fn{
\lf.{{\p^2\Delta_A}\o{\p V\p \ov V}}\ri|_{V=0}=
{{2b_{A}^{N=2}}\o{(-i T+i \ov T)(-i U+i \ov U)}}\ ,}
verifying the integrability condition of \agntz.

\appendix{\appD}{$\IZ_6$--orbifold limits of $K3$ and their instanton numbers}


\def\zero{\ss{E_8}}

\def\two{\ss{U(1)\times E_7}}
\def\three{\ss{U(1)\times E_7}}
\def\four{\ss{SU(3)\times E_6}}
\def\five{\ss{U(1)\times SU(2)\times E_6}}
\def\six{\ss{U(1)^2\times E_6}}
\def\seven{\ss{SO(16)}}
\def\eight{\ss{U(1)\times SO(14)}}

\def\ten{\ss{U(1)\times SU(2)\times SO(12)}}
\def\eleven{\ss{U(1)\times SU(2)\times SO(12)}}
\def\twelve{\ss{U(1)^2\times SO(12)}}
\def\thirten{\ss{U(1)\times SU(3)\times SO(10)}}
\def\fourten{\ss{U(1)\times SU(2)^2\times SO(10)}}
\def\fivten{\ss{U(1)^2\times SU(2)\times SO(10)}}
\def\sixten{\ss{U(1)\times SU(4)\times SO(8)}}
\def\seventen{\ss{SU(9)}}
\def\eighten{\ss{U(1)\times SU(8)}}\
\def\nineten{\ss{U(1)\times SU(8)}}
\def\twenty{\ss{U(1)\times SU(2)\times SU(7)}}
\def\twentyone{\ss{U(1)^2\times SU(7)}}
\def\twentytwo{\ss{U(1)^2\times SU(7)}}
\def\twentythree{\ss{SU(2)\times SU(3)\times SU(6)}}
\def\twentyfour{\ss{U(1)\times SU(3)\times SU(6)}}
\def\twentyfive{\ss{U(1)\times SU(2)^2\times SU(6)}}
\def\twentysix{\ss{U(1)\times SU(4)\times SU(5)}}

\def\zeros{\ss{E_8'}}
\def\ones{\ss{SU(2)'\times E_7'}}
\def\twos{\ss{U(1)'\times E_7'}}
\def\threes{\ss{U(1)'\times E_7'}}
\def\fours{\ss{SU(3)'\times E_6'}}
\def\fives{\ss{U(1)'\times SU(2)'\times E_6'}}
\def\sixs{\ss{U(1)'^2\times E_6'}}
\def\sevens{\ss{SO(16)'}}
\def\eights{\ss{U(1)'\times SO(14)'}}
\def\nines{\ss{U(1)'\times SO(14)'}}
\def\tens{\ss{U(1)'\times SU(2)'\times SO(12)'}}
\def\elevens{\ss{U(1)'\times SU(2)'\times SO(12)'}}
\def\twelves{\ss{U(1)'^2\times SO(12)'}}
\def\thirtens{\ss{U(1)'\times SU(3)'\times SO(10)'}}
\def\fourtens{\ss{U(1)'\times SU(2)'^2\times SO(10)'}}
\def\fivtens{\ss{U(1)'^2\times SU(2)'\times SO(10)'}}
\def\sixtens{\ss{U(1)'\times SU(4)'\times SO(8)'}}

\def\twentythrees{\ss{SU(2)'\times SU(3)'\times SU(6)'}}

\def\twentyfives{\ss{U(1)'\times SU(2)'^2\times SU(6)'}}

{\vbox{\ninepoint{
\def\ss#1{{\scriptstyle{#1}}}
$$
\vbox{\offinterlineskip\tabskip=0pt
\halign{\strut\vrule#
&~$#$~\hfil
&~$#$~\hfil
&\vrule#
&~$#$~\hfil
&\vrule#
&~$#$~\hfil
&\vrule#
&~$#$~\hfil
&\vrule#
\cr
\noalign{\hrule}
&
\ss{\gamma}
&
\ss{\tilde\gamma}
&&
\ss{\rm \ perturbative\ gauge\ group}
&&
\ss{(n^{(1)},n^{(2)})}
&
\cr
\noalign{\hrule}
\noalign{\hrule}
&
\ss{(1,1,0,0,0,0,0,0)}
&
\ss{(0,0,0,0,0,0,0,0)}
&&
\two \times \zeros
&&
\ss{(24,0)}
&
\cr
&
\ss{(3,2,1,0,0,0,0,0)}
&
\ss{(0,0,0,0,0,0,0,0)}
&&
\six \times \zeros
&&
\ss{(24,0)}
&
\cr
&
\ss{(5,1,0,0,0,0,0,0)}
&
\ss{(0,0,0,0,0,0,0,0)}
&&
\eleven\times\zeros
&&
\ss{(24,0)}
&
\cr
&
\ss{\h(7,1,1,1,1,1,1,-1)}
&
\ss{(0,0,0,0,0,0,0,0)}
&&
\twenty\times\zeros
&&
\ss{(24,0)}
&
\cr
&
\ss{\h(9,3,3,1,1,1,1,1)}
&
\ss{(0,0,0,0,0,0,0,0)}
&&
\twentyfour\times\zeros
&&
\ss{(24,0)}
&
\cr
&
\ss{(2,2,0,0,0,0,0,0)}
&
\ss{(3,3,0,0,0,0,0,0)}
&&
\three\times\ones
&&
\ss{(12,12)}
&
\cr
&
\ss{(4,2,0,0,0,0,0,0)}
&
\ss{(3,3,0,0,0,0,0,0)}
&&
\ten\times\ones
&&
\ss{(20,4)}
&
\cr
&
\ss{(5,1,1,1,1,1,1,1)}
&
\ss{(3,3,0,0,0,0,0,0)}
&&
\seventen\times\ones
&&
\ss{(14,10)}
&
\cr
&
\ss{(1,1,1,1,1,1,1,-1)}
&
\ss{(3,3,0,0,0,0,0,0)}
&&
\eighten\times\ones
&&
\ss{(18,6)}
&
\cr
&
\ss{(5,1,1,1,1,1,1,-1)}
&
\ss{(3,3,0,0,0,0,0,0)}
&&
\nineten\times\ones
&&
\ss{(20,4)}
&
\cr
\noalign{\hrule}
&
\ss{(2,2,2,2,2,0,0,0)}
&
\ss{(3,3,0,0,0,0,0,0)}
&&
\twentysix\times\ones
&&
\ss{(20,4)}
&
\cr
&
\ss{(4,2,2,0,0,0,0,0)}
&
\ss{(1,1,0,0,0,0,0,0)}
&&
\four\times\twos
&&
\ss{(12,12)}
&
\cr
&
\ss{(6,0,0,0,0,0,0,0)}
&
\ss{(1,1,0,0,0,0,0,0)}
&&
\seven\times\twos
&&
\ss{(8,16)}
&
\cr
&
\ss{(2,2,2,0,0,0,0,0)}
&
\ss{(1,1,0,0,0,0,0,0)}
&&
\thirten\times\twos
&&
\ss{(18,6)}
&
\cr
&
\ss{(3,3,1,1,1,1,1,-1)}
&
\ss{(1,1,0,0,0,0,0,0)}
&&
\twentyfive\times\twos
&&
\ss{(18,6)}
&
\cr
&
\ss{(2,1,1,0,0,0,0,0)}
&
\ss{(2,2,0,0,0,0,0,0)}
&&
\five\times\threes
&&
\ss{(18,6)}
&
\cr
&
\ss{(4,1,1,0,0,0,0,0)}
&
\ss{(2,2,0,0,0,0,0,0)}
&&
\fivten\times\threes
&&
\ss{(4,20)}
&
\cr
&
\ss{(5,1,1,1,1,1,0,0)}
&
\ss{(2,2,0,0,0,0,0,0)}
&&
\twentythree\times\threes
&&
\ss{(4,20)}
&
\cr
&
\ss{(3,2,1,0,0,0,0,0)}
&
\ss{(4,2,2,0,0,0,0,0)}
&&
\six\times\fours
&&
\ss{(12,12)}
&
\cr
&
\ss{(5,1,0,0,0,0,0,0)}
&
\ss{(4,2,2,0,0,0,0,0)}
&&
\eleven\times\fours
&&
\ss{(14,10)}
&
\cr
\noalign{\hrule}
&
\ss{\h(7,1,1,1,1,1,1,-1)}
&
\ss{(4,2,2,0,0,0,0,0)}
&&
\twenty\times\fours
&&
\ss{(17,7)}
&
\cr
&
\ss{\h(9,3,3,1,1,1,1,1)}
&
\ss{(4,2,2,0,0,0,0,0)}
&&
\twentyfour\times\fours
&&
\ss{(18,6)}
&
\cr
&
\ss{(4,2,0,0,0,0,0,0)}
&
\ss{(2,1,1,0,0,0,0,0)}
&&
\ten\times\fives
&&
\ss{(12,12)}
&
\cr
&
\ss{(5,1,1,1,1,1,1,1)}
&
\ss{(2,1,1,0,0,0,0,0)}
&&
\seventen\times \fives
&&
\ss{(11,13)}
&
\cr
&
\ss{(1,1,1,1,1,1,1,-1)}
&
\ss{(2,1,1,0,0,0,0,0)}
&&
\eighten\times\fives
&&
\ss{(15,9)}
&
\cr
&
\ss{(5,1,1,1,1,1,1,-1)}
&
\ss{(2,1,1,0,0,0,0,0)}
&&
\nineten\times\fives
&&
\ss{(11,13)}
&
\cr
&
\ss{(2,2,2,2,2,0,0,0)}
&
\ss{(2,1,1,0,0,0,0,0)}
&&
\twentysix\times \fives
&&
\ss{(16,8)}
&
\cr
&
\ss{(6,0,0,0,0,0,0,0)}
&
\ss{(3,2,1,0,0,0,0,0)}
&&
\seven \times \sixs
&&
\ss{(8,16)}
&
\cr
&
\ss{(2,2,2,0,0,0,0,0)}
&
\ss{(3,2,1,0,0,0,0,0)}
&&
\thirten\times \sixs
&&
\ss{(16,8)}
&
\cr
&
\ss{(3,3,1,1,1,1,1,-1)}
&
\ss{(3,2,1,0,0,0,0,0)}
&&
\twentyfive\times \sixs
&&
\ss{(17,7)}
&
\cr
&
\ss{(5,1,0,0,0,0,0,0)}
&
\ss{(6,0,0,0,0,0,0,0)}
&&
\eleven\times \sevens
&&
\ss{(16,8)}
&
\cr
\noalign{\hrule}}
\hrule}$$
\vskip-10pt
\vskip10pt}}}


{\vbox{\ninepoint{
\def\ss#1{{\scriptstyle{#1}}}
$$
\vbox{\offinterlineskip\tabskip=0pt
\halign{\strut\vrule#
&~$#$~\hfil
&~$#$~\hfil
&\vrule#
&~$#$~\hfil
&\vrule#
&~$#$~\hfil
&\vrule#
&~$#$~\hfil
&\vrule#
\cr
\noalign{\hrule}
&
\ss{\gamma}
&
\ss{\tilde\gamma}
&&
\ss{\rm \ perturbative\ gauge\ group}
&&
\ss{(n^{(1)},n^{(2)})}
&
\cr
\noalign{\hrule}
\noalign{\hrule}
&
\ss{\h(7,1,1,1,1,1,1,-1)}
&
\ss{(6,0,0,0,0,0,0,0)}
&&
\twenty\times \sevens
&&
\ss{(16,8)}
&
\cr
&
\ss{\h(9,3,3,1,1,1,1,1)}
&
\ss{(6,0,0,0,0,0,0,0)}
&&
\twentyfour\times \sevens
&&
\ss{(16,8)}
&
\cr
&
\ss{(3,1,0,0,0,0,0,0)}
&
\ss{(2,0,0,0,0,0,0,0)}
&&
\twelve\times \eights
&&
\ss{(12,12)}
&
\cr
&
\ss{(3,3,2,0,0,0,0,0)}
&
\ss{(2,0,0,0,0,0,0,0)}
&&
\fourten\times \eights
&&
\ss{(12,12)}
&
\cr
&
\ss{\h(9,1,1,1,1,1,1,1)}
&
\ss{(2,0,0,0,0,0,0,0)}
&&
\twentytwo\times \eights
&&
\ss{(12,12)}
&
\cr
&
\ss{(3,1,0,0,0,0,0,0)}
&
\ss{(4,0,0,0,0,0,0,0)}
&&
\twelve\times\nines
&&
\ss{(16,8)}
&
\cr
&
\ss{(3,3,2,0,0,0,0,0)}
&
\ss{(4,0,0,0,0,0,0,0)}
&&
\fourten\times\nines
&&
\ss{(16,8)}
&
\cr
&
\ss{\h(9,1,1,1,1,1,1,1)}
&
\ss{(4,0,0,0,0,0,0,0)}
&&
\twentytwo\times\nines
&&
\ss{(16,8)}
&
\cr
&
\ss{(4,1,1,0,0,0,0,0)}
&
\ss{(4,2,0,0,0,0,0,0)}
&&
\fivten\times\tens
&&
\ss{(14,10)}
&
\cr
&
\ss{(5,1,1,1,1,1,0,0)}
&
\ss{(4,2,0,0,0,0,0,0)}
&&
\twentythree\times\tens
&&
\ss{(14,10)}
&
\cr
\noalign{\hrule}
&
\ss{(2,2,2,0,0,0,0,0)}
&
\ss{(5,1,0,0,0,0,0,0)}
&&
\thirten\times\elevens
&&
\ss{(16,8)}
&
\cr
&
\ss{(3,3,1,1,1,1,1,-1)}
&
\ss{(5,1,0,0,0,0,0,0)}
&&
\twentyfive\times\elevens
&&
\ss{(16,8)}
&
\cr
&
\ss{(5,1,1,1,0,0,0,0)}
&
\ss{(3,1,0,0,0,0,0,0)}
&&
\sixten\times \twelves
&&
\ss{(14,10)}
&
\cr
&
\ss{(3,1,1,1,1,1,1,1)}
&
\ss{(3,1,0,0,0,0,0,0)}
&&
\twentyone\times \twelves
&&
\ss{(14,10)}
&
\cr
&
\ss{\h(7,1,1,1,1,1,1,-1)}
&
\ss{(2,2,2,0,0,0,0,0)}
&&
\twenty\times \thirtens
&&
\ss{(12,12)}
&
\cr
&
\ss{\h(9,3,3,1,1,1,1,1)}
&
\ss{(2,2,2,0,0,0,0,0)}
&&
\twentyfour\times \thirtens
&&
\ss{(13,11)}
&
\cr
&
\ss{(5,1,1,1,0,0,0,0)}
&
\ss{(3,3,2,0,0,0,0,0)}
&&
\sixten\times \fourtens
&&
\ss{(14,10)}
&
\cr
&
\ss{(3,1,1,1,1,1,1,1)}
&
\ss{(3,3,2,0,0,0,0,0)}
&&
\twentyone\times\fourtens
&&
\ss{(13,11)}
&
\cr
&
\ss{(5,1,1,1,1,1,1,1)}
&
\ss{(4,1,1,0,0,0,0,0)}
&&
\seventen\times \fivtens
&&
\ss{(10,14)}
&
\cr
&
\ss{(1,1,1,1,1,1,1,-1)}
&
\ss{(4,1,1,0,0,0,0,0)}
&&
\eighten \times \fivtens
&&
\ss{(14,10)}
&
\cr
\noalign{\hrule}
&
\ss{(5,1,1,1,1,1,1,-1)}
&
\ss{(4,1,1,0,0,0,0,0)}
&&
\nineten\times \fivtens
&&
\ss{(11,13)}
&
\cr
&
\ss{(2,2,2,2,2,0,0,0)}
&
\ss{(4,1,1,0,0,0,0,0)}
&&
\twentysix\times \fivtens
&&
\ss{(15,9)}
&
\cr
&
\ss{\h(9,1,1,1,1,1,1,1)}
&
\ss{(5,1,1,1,0,0,0,0)}
&&
\twentytwo\times \sixtens
&&
\ss{(11,13)}
&
\cr
&
\ss{(5,1,1,1,1,1,1,1)}
&
\ss{(5,1,1,1,1,1,0,0)}
&&
\seventen\times \twentythrees
&&
\ss{(10,12)}
&
\cr
&
\ss{(1,1,1,1,1,1,1,-1)}
&
\ss{(5,1,1,1,1,1,0,0)}
&&
\eighten\times \twentythrees
&&
\ss{(4,20)}
&
\cr
&
\ss{(5,1,1,1,1,1,1,-1)}
&
\ss{(5,1,1,1,1,1,0,0)}
&&
\nineten\times \twentythrees
&&
\ss{(1,23)}
&
\cr
&
\ss{(2,2,2,2,2,0,0,0)}
&
\ss{(5,1,1,1,1,1,0,0)}
&&
\twentysix\times \twentythrees
&&
\ss{(4,20)}
&
\cr
&
\ss{\h(7,1,1,1,1,1,1,-1)}
&
\ss{(3,3,1,1,1,1,1,-1)}
&&
\twenty\times \twentyfives
&&
\ss{(12,12)}
&
\cr
&
\ss{\h(9,3,3,1,1,1,1,1)}
&
\ss{(3,3,1,1,1,1,1,-1)}
&&
\twentyfour\times\twentyfives
&&
\ss{(12,12)}
&
\cr
\noalign{\hrule}}
\hrule}$$
\vskip-10pt
\noindent{\bf Table 8:}
{\sl All 60 $T^4/\IZ_6$--orbifolds with gauge twist 
$(\gamma,\tilde\gamma)$:
Their perturbative gauge group and instanton numbers
$(n^{(1)},n^{(2)})$ w.r.t. $SU(2)$--bundles.}
\vskip10pt}}}


\listrefs
\end